\documentclass[trackchanges,preprint2]{aastex}
\turnoffedit
\usepackage{hyperref}
\usepackage{lineno}
\usepackage{tabularx}
\modulolinenumbers[5]
\usepackage{url}
\usepackage{color}
\usepackage{graphicx}
\usepackage{epstopdf}
\usepackage{array}
\newcolumntype{L}[1]{>{\raggedright\let\newline\\\arraybackslash\hspace{0pt}}m{#1}}
\usepackage{footnote}
\makesavenoteenv{tabularx}
\makesavenoteenv{table}

\shorttitle{gPhoton - Tools for GALEX Photon Events}
\shortauthors{Chase Million et al.}

\begin{document}
\title{gPhoton: The GALEX Photon Data Archive}

\author{
  Chase Million\altaffilmark{1},
  Scott W. Fleming\altaffilmark{2,3},
  Bernie Shiao\altaffilmark{2},
  Mark Seibert\altaffilmark{4},
  Parke Loyd\altaffilmark{5},
  Michael Tucker\altaffilmark{6},
  Myron Smith\altaffilmark{2,7},
  Randy Thompson\altaffilmark{2,3},
  Richard L. White\altaffilmark{2}
}

\altaffiltext{1}{Million Concepts LLC, PO Box 119, 141 Mary St, Lemont, PA 16851, USA}
\altaffiltext{2}{Space Telescope Science Institute, 3700 San Martin Dr, Baltimore, MD 21218, USA}
\altaffiltext{3}{CSRA, Inc., 3700 San Martin Dr, Baltimore, MD 21218, USA}
\altaffiltext{4}{The Observatories of the Carnegie Institution of Washington, 813 Santa Barbara Street, Pasadena, CA 91101, USA}
\altaffiltext{5}{Laboratory for Atmospheric and Space Physics, Boulder, Colorado,  80309}
\altaffiltext{6}{Dept. of Physics and Astronomy, Appalachian State University, Boone, NC 28608, USA}
\altaffiltext{7}{Current Address: National Optical Astronomy Observatory, 950 N. Cherry Ave., Tucson, AZ 85719}


\begin{abstract}
gPhoton is a new database product and software package that enables analysis of GALEX ultraviolet data at the photon level. The project's stand-alone, pure-Python calibration pipeline reproduces the functionality of the original mission pipeline to reduce raw spacecraft data to lists of time-tagged, sky-projected photons, which are then hosted in a publicly available database by the Mikulski Archive at Space Telescope (MAST). This database contains approximately 130 terabytes of data describing approximately 1.1 trillion sky-projected events with a timestamp resolution of five milliseconds. A handful of Python and command line modules serve as a front-end to interact with the database and to generate calibrated light curves and images from the photon-level data at user-defined temporal and spatial scales. The gPhoton software and source code are in active development and publicly available under a permissive license. We describe the motivation, design, and implementation of the calibration pipeline, database, and tools, with emphasis on divergence from prior work, as well as challenges created by the large data volume. We summarize the astrometric and photometric performance of gPhoton relative to the original mission pipeline. For a brief example of short time domain science capabilities enabled by gPhoton, we show new flares from the known M dwarf flare star CR Draconis. The gPhoton software has permanent object identifiers with the ASCL (ascl:1603.004) and DOI (doi:10.17909/T9CC7G). \edit1{This paper describes the software as of version v1.27.2.}
\end{abstract}

\section{GALEX Overview}
The Galaxy Evolution Explorer \citep{mar2005} was a NASA Small Explorer (SMEX) telescope that surveyed the sky in the ultraviolet over ten years between launch on 28 April 2003 and spacecraft termination on 28 June 2013. The spacecraft, instruments, data, and calibration are well described in previous publications \citep{mor2005,mor2007} and the mission`s online technical documentation.\footnote{\url{http://www.galex.caltech.edu/wiki/Public:Documentation}} We will restrict discussion to topics that are necessary for completeness, have not appeared elsewhere in the literature, or are of particular importance to the gPhoton project.

GALEX carried two micro-channel plate detectors (MCP) with 1.25 degree fields-of-view (FoV), simultaneously exposed via a dichroic. \edit1{The detectors record signals from electrical cascades, referred to as ``events,'' which were produced by photons hitting the MCPs. Detector positions and time stamps of these events recorded by the spacecraft were then corrected for instrumental effects} and re-projected into celestial coordinates by a calibration pipeline on the ground. The detectors observed in two broad ultraviolet (UV) bands centered around $1528\,\rm{\AA}$ (Far Ultraviolet or ``FUV'') and $2271\,\rm{\AA}$ (Near Ultraviolet or ``NUV''). The FUV detector failed in May of 2009, but the NUV detector continued to operate until the end of the mission. The spacecraft could observe in either direct imaging or slitless spectroscopic (grism) modes. Observations were conducted while the spacecraft was on the night side of each orbit (an ``eclipse''), which lasted 1500-1800 seconds. To avoid detector burn-in or local gain sag effects caused by depletion of electrons in the multiplier plate, the telescope did not stare at a fixed location on the sky during an observation but continuously moved the boresight relative to the target position. Several boresight patterns, or ``modes,'' were used over the course of the mission, which impacted the nature of the corresponding observational data.

In the most basic ``dither'' mode, the spacecraft boresight would trace out a tight spiral pattern with a radius of $\sim1'$. Dither mode was used most often for Deep or Medium Imaging Surveys (DIS, MIS) in which a full eclipse of $\sim1600$ seconds was spent observing a single region of the sky. In the All-sky Imaging Survey (AIS) mode, the spacecraft boresight would jump between multiple positions (or ``legs'') on the sky for short integrations of $\sim100$ seconds each. Between each leg, the detector was set to a non-observing, low voltage state. This resulted in one independent observation (or ``visit'') per leg. Another mode, called ``petal pattern,'' was used to distribute the flux from particularly bright targets across the detector. Petal pattern is in some ways similar to the AIS mode, but the legs were tightly clustered into the approximate area of a single FoV and the detector remained in its nominal high-voltage state in between.

On 4 May 2010, the ``Coarse Sun Point'' (CSP) anomaly---\edit1{a reference} to the safe mode entered by the spacecraft at that time---resulted in image degradation of the NUV detector. The CSP anomaly precipitated severe streaking in the detector's Y-direction, likely due to a failed capacitor. Although the effect was largely corrected through subsequent calibration and on-board adjustments, observations taken between 4 May and 23 June 2010 have substantially worse point spread functions (PSF). Care should be used when comparing observations made before this time range to observations made after to discount bias due to either degraded PSF or uncorrected ``ghost'' photons.\footnote{\url{http://www.galex.caltech.edu/wiki/Public:Documentation/Chapter_8}}

NASA support for the mission ended in February of 2011. At that time, ownership of the spacecraft was transferred to the California Institute of Technology for a phase called the ``Complete the All-sky UV Survey Extension'' (CAUSE), during which operating costs were solicited from individuals or institutions, and spacecraft engineering constraints related to field and source brightness were relaxed, making it possible to observe bright regions of the sky that were off limits during the primary mission.\footnote{\url{http://www.galex.caltech.edu/cause/index.html}} Spacecraft slew rate limits were also relaxed, permitting a high-coverage ``scan mode'' that swept across several degrees of sky in a single integration \edit1{\citep{olmedo2015deep}}. Ownership of the CAUSE-phase data resides with each of the primary investigators, and only a small fraction of it has been made available to the public through MAST at the time of writing.  Although the new calibration capabilities described herein may be of particular value in using and interpreting CAUSE data generally, and scan mode observations of very bright or dense fields in particular, this paper \edit1{and the current gPhoton database only cover} the direct imaging data through the end of the NASA-supported mission, corresponding to General Release 7 (GR7) in the MAST archives. Through GR7, GALEX collected data over 34,389 direct image eclipses, covering $\sim76.9\%$ of the sky in at least one band. Future work may add gPhoton support for CAUSE phase, scan mode, or spectroscopic data collected throughout the mission.

In Section \ref{motivation} we describe the motivation behind constructing the gPhoton database and software suite. In Section \ref{database} we describe the design and content of the $\sim 1.1$ trillion row database hosted at MAST. In Section \ref{softwaretools} we describe the primary modules for generating photon lists, light curves and images. In Section \ref{calibration} we present tests of the calibration precision with respect to astrometry \edit1{and} photometry in relation to the mission catalogs, and photometry in relation to a calibration standard. In Section \ref{implementation} we discuss implementation challenges and solutions. Finally, in Section \ref{scienceexamples}, we highlight an example science case enabled by gPhoton: stellar flares of CR Draconis.

\edit1{This paper describes version 1.27.2 of gPhoton. The software is under active development, and users are encouraged to consult the online documentation to supplement the information presented herein.}

\section{Motivation}
\label{motivation}
Micro-channel plate (MCP) detectors like those on GALEX are non-integrating imagers---sometimes called ``photon-counting''---that can record position and time information individually for \edit1{each detected electrical cascade ``event.'' The majority of such events are initiated by photons with astrophysical origins, but they may also be due to instrument noise or artificial ``stims'' used for calibration. Some number of photons interact with the detector but are not recorded as events (e.g., because of dead time as described in Section \ref{deadtimedesc}).} The GALEX detectors were capable of recording data with a time resolution of five microseconds, though the vast majority of observations were made in a compressed mode at five millisecond resolution. Due primarily to computer storage and processing constraints, calibrated GALEX data were only released and archived by the mission as either per-observation or multi-observation (coadded) image maps with exposure depths on the order of hundreds to thousands of seconds. Although the GALEX mission's data calibration pipeline (hereafter referred to as the ``mission pipeline'') was capable of producing aspect-corrected photon list files (called ``extended'' or ``x-files''), this was rarely done only as part of instrument diagnostics or by special request of members of the scientific community \citep{rob2005, wel2006, wel2007}. The team produced very little documentation about the detector performance or calibration on timescales shorter than $100$ seconds.

Advances in data storage and processing capabilities now make archiving, distribution, and analysis of the photon-level data technologically feasible. By the end of the mission, however, the mission pipeline had grown to sufficient complexity and dependence on its software and hardware operating environment that attempts to run it outside of the networked infrastructure upon which it was developed at Caltech proved unsuccessful. We undertook the gPhoton project,\footnote{\url{http://dx.doi.org/doi:10.17909/T9CC7G} \textbar \url{http://ascl.net/1603.004} \textbar \url{https://github.com/cmillion/gPhoton}} in part, to migrate key functionality of the mission pipeline into a stand-alone, open source software base that is robust enough to operating environment to serve as a jumping off point for future researchers to modify or improve the calibration or otherwise build on the legacy of this unique data set. Another major objective was to enable the creation of calibrated light curves and images at user-specified spatial and temporal scales, permitting studies of short time-domain variability in the ultraviolet over a significant fraction of the sky for the first time. The gPhoton project design goals included the following key features:
\begin{itemize}
\item{A stand-alone GALEX calibration pipeline that reproduces the capabilities of the mission pipeline to reduce spacecraft data to time-tagged, aspect-corrected photon event data.}
\item{A publicly accessible database containing nearly all photon events from the mission.}
\item{Software that can perform necessary scientific calibrations (astrometric, photometric, exposure time, etc.), at quality comparable to the original mission pipeline over visit-level timescales.}
\item{An ability to flexibly create images (as a coadd over one or more specified time ranges) or image cubes (as sequences of such coadds).}
\item{An ability to create light curves with custom binning.}
\item{Lower the barrier to entry of working with short time domain GALEX data by, e.g., minimizing the number of primary (forward-facing) modules required, wrapping the database queries behind a Python interface, and using widely supported output file formats like FITS and comma separated values (CSV).}
\end{itemize}

While the gPhoton project does reproduce much of the core functionality of the mission pipeline, it is not intended as either a full migration or a faithful port of the original mission pipeline. As will be described, some archived output files from the mission pipeline are used as inputs where deemed expedient, and the calibration and reduction methodology has been modified in places in service to both computational efficiency and the unique properties and uses of photon-level data. The gPhoton tools also do not include a capability for automated source \emph{detection} (i.e. catalog creation).

\section{The Database}
\label{database}
\subsection{\edit1{Mission Pipeline Data Products}}
During the GALEX mission, data were downlinked from the spacecraft and assembled on the ground into monolithic telemetry files (-tlm). The ``ingest'' stage of the mission pipeline split these into various types of encoded raw detector event and spacecraft state (-scst) data, which included coarse aspect solutions from the on-board star tracker at one-second resolution, as well as spacecraft housekeeping records. The most important class of encoded raw detector data for gPhoton, containing nominal scientific observations, were the -raw6 files. \edit1{The aspect solution was used to translate the photon data from the spacecraft reference frame onto the celestial sphere to create images, and, a refined aspect solution (-asprta) was generated by iteratively comparing sequences of such images to star catalogs.}

\subsection{\edit1{Reduction of the Photon Data By gPhoton}}
The -raw6 are decoded with a sequence of bitwise manipulations into lists of raw detector positions ($x$ and $y$) with timestamps for all detector events and further adjusted with ``static'' (in the detector reference frame) calibrations for wiggle, walk, nonlinearity and distortion, all described more completely in \citet{mor2007}. For post-CSP data (after eclipse number 37460), the detector calibration was modified to correct and account for changes in the detector hardware and software. The most substantial of the post-CSP calibration changes was the addition of a processing step to correct for detector streaking caused by the anomaly, correlated strongly to the YA value of the raw position data (one of many intermediate raw data values used in derivation of detector event positions, described in Table 2 of \citet{mor2007}). \edit1{The gPhoton software then uses the refined spacecraft attitude (-asprta) and spacecraft state (-scst) files to compute the celestial coordinates of photon events, which are exported to CSV files.} These files contain the timestamps of photon event, the event positions on the detector, the aspect-corrected positions on the sky (\edit2{as right ascension and declination}), and status flags used to track a variety of conditions related to the detector readout. The vast majority of users will only be interested in photon events for which the photon file flag value is equal to zero, indicating nominally processed data.

A subset of data are not aspect-correctable because they fall in time ranges that are not covered by the refined aspect solutions. Such gaps occurred when the detector voltage was ramping up or down between observations, the slew rate was too high (as between legs of a petal pattern observation), or the stellar field was too sparse for the aspect refinement pipeline to obtain a solution. Some events, associated with electrical pulsers used for calibration, detector noise, or downlink errors, also cannot be aspect corrected because they fall outside the detector FoV. Therefore, four CSV files are created: one file for each band that contains aspect-corrected photon events, and one file for each band that contains photon events that were not aspect-corrected. The non-aspect-corrected photon events are retained and loaded into a database to be used for estimating dead time corrections (per Section \ref{deadtimedesc}). When events cannot be aspect-corrected, their right ascension and declination values are assigned values of \emph{NULL} in the photon list file; for this reason, we refer to uncorrected ``null data'' and nominal ``non-null data.''

\subsection{Database Structure}
For performance optimization purposes, the event-level data is partitioned in the following manner.  The smallest unit of partitioning is called a ``zone,'' which has a fixed height of $30''$ in declination.  A varying number of zones are further grouped into ``partitions,'' where each partition stores approximately the same number of photon events. The gPhoton project uses a total of ten databases, each having a separate table for FUV and NUV, with varying numbers of partitions.  The number of partitions per database is assigned such that the total number of rows per table per database is approximately the same. All together, there are a total of 21600 zones divided across 999 partitions.

We make use of the fast zone matching algorithm described in \citet{gra2006} for loading and querying the database. Both the database boundaries and the number of $30''$ zones assigned to each partition were defined using an assumption that the total number of photon events in a given eclipse is distributed evenly across that eclipse's footprint. The cross-section of each eclipse's footprint against the zone boundaries is calculated to determine which zones that eclipse overlaps. The number of photons in each zone from this eclipse is estimated based on the cross-sectional area, e.g., if a given eclipse spans two zones, but only 10\% of the eclipse's footprint is in one of the zones, 90\% of its total photon events would be considered to belong to the first zone, and 10\% to the other. This allowed us to assign zones to each partition without the need to calculate the zone assignment of all the photon events ahead of time. When the databases were actually populated, the zone assignment for each photon event was calculated individually.

\begin{figure}[h!]
\includegraphics[width=0.46\textwidth,keepaspectratio]{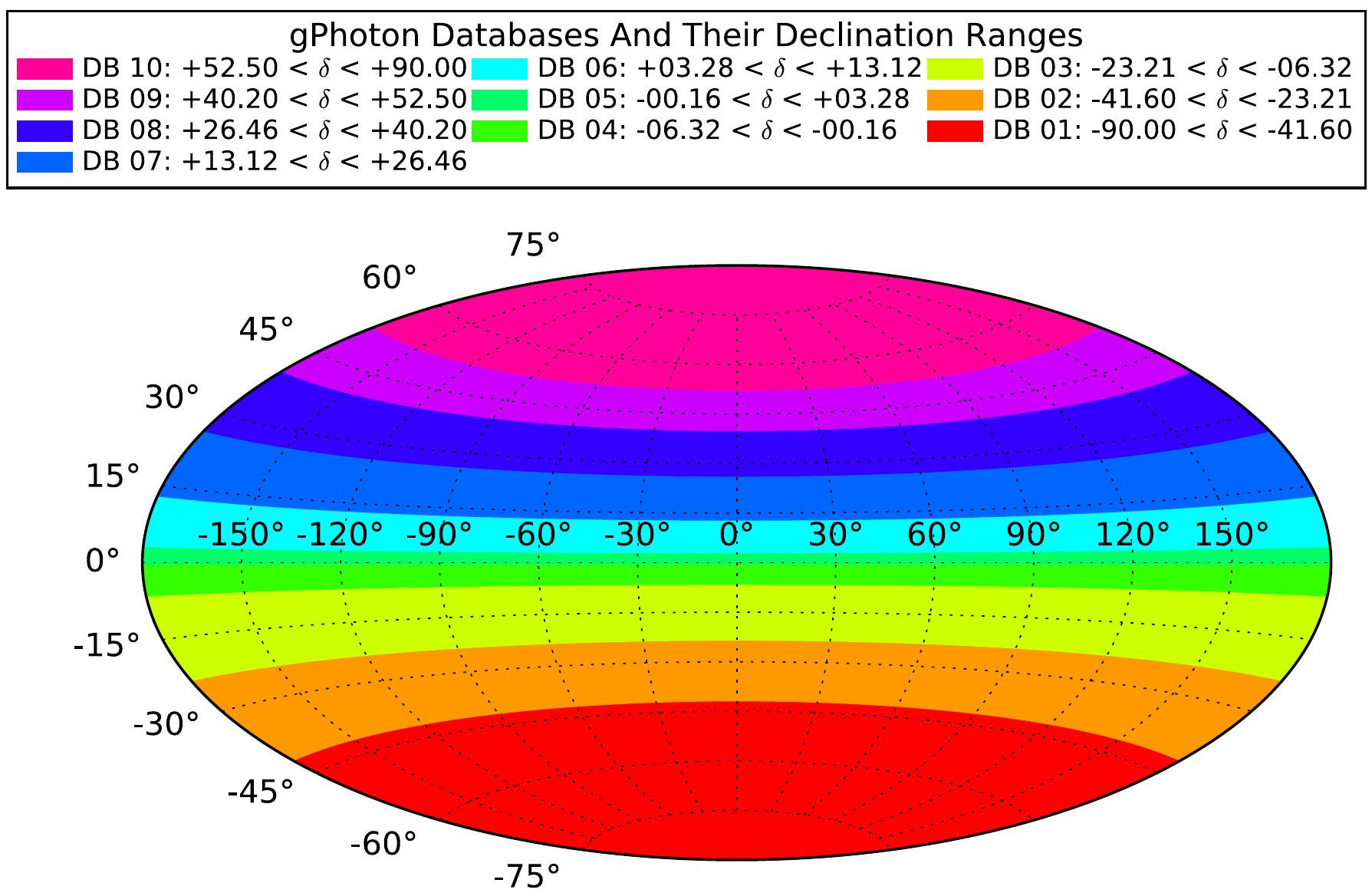}
\caption{The declination boundaries of the ten individual databases holding the full corpus of photon data. There are a total of 999 partitions across the ten databases, each with a variable number of $30''$ zones (stripes of declination).  The number of zones in each partition, and the number of partitions in each database, were assigned so that the size of the ten databases would be roughly equal to each other. \label{dbdist}}
\end{figure}

The distribution of the ten databases on the sky is shown in Figure \ref{dbdist}, along with a table summarizing the declination ranges and number of photon events in each database (Table \ref{dbcounts}). Given the possibility that a query could span two or more databases, events are assigned to one and only one database. The majority of normal queries access only a single database, but those that do span more are handled on the server, transparent to both the software and end users.
	NULL data reside in a single database that is partitioned and indexed on photon event time. To optimize for common classes of queries, the non-NULL data are indexed in three ways:
\begin{enumerate}
	\item{\emph{zoneID} and photon event \emph{time} - Sky coordinates, a search radius, and a time range are inputs to a number of database cone search functions. The declination and radius are translated into a range of zoneIDs, which form the basis for construction of an SQL query.}
	\item{\emph{zoneID}, \emph{RA}, and \emph{Dec} - Often used for sub-queries in the functions described above, and occasionally used in queries where photon event time is not a parameter.}
	\item{Photon even \emph{time} and \emph{flag} value - To optimize queries based on time range alone.}
\end{enumerate}

\begin{table}[h!]
\caption{Distribution of non-null photon events per database.}
\begin{tabularx}{0.45\textwidth}{lllll}
\hline
DB & min $\delta$ & max $\delta$ & N\_FUV & N\_NUV\\
   & deg          & deg          & x$10^9$ & x$10^9$\\
\hline
1 & -90.00 & -41.60 &   6.257390699 &  86.634587040\\
2 & -41.60 & -23.21 &   6.381213422 &  86.548536265\\
3 & -23.21 &  -6.32 &   6.581269082 &  86.661438444\\
4 &  -6.32 &  -0.16 &   5.855815141 &  86.824342990\\
5 &  -0.16 &   3.28 &   5.918115466 &  87.675967291\\
6 &   3.28 &  13.12 &   6.431868719 &  86.895361749\\
7 &  13.12 &  26.46 &   6.230197521 &  87.197846131\\
8 &  26.46 &  40.20 &   5.392882518 &  86.855190790\\
9 &  40.20 &  52.50 &   5.394019934 &  86.683794328\\
10 &  52.50 &  90.00 &   6.540819967 & 100.482179843\\
\hline
\end{tabularx}
\label{dbcounts}
\end{table}

\section{The Software Tools}
\label{softwaretools}
There are four primary modules included in gPhoton and described in Table \ref{moduledesc}. These utilities are all written in Python and released under a permissive license. With the exception of gPipeline, the tools can be called either from the command line or imported as Python modules. When imported as modules, output is also returned as Python objects that include the complete lists of photon events used. The command line utilities draw upon a large number of supporting functions which will not be described in this paper, but are possibly of interest to users who want to perform advanced or specialized analyses with the gPhoton data, or even modify the functionality to fit their individual needs. For more information, users are encouraged to consult the documentation available in the software repository, the User Guide,\footnote{\url{https://github.com/cmillion/gPhoton/blob/master/docs/UserGuide.md}} or the MAST page for the project.\footnote{\url{https://archive.stsci.edu/prepds/gphoton/}}

\begin{table}[htbp!]
\caption{Summary of Primary gPhoton Modules with example Python syntax.}
\begin{tabularx}{0.45\textwidth}{|L{0.095\textwidth}|L{0.315\textwidth}|}
\hline
	{\bf Module} & {\bf Function}\\\hline
	gPipeline & Generates aspect-corrected photon lists from a small set of user supplied input files. Output from gPipeline was used to populate the photon event database that the other modules query and, therefore, the majority of researchers will not need this module. \edit1{Please see the User Guide for syntax.}\\\hline
	gFind & Provides information on the available raw exposure depths and time ranges for any location on the sky. \edit1{Example to find the total available exposure depth, in each band, for the star CR Draconis: \parbox[t]{0.315\textwidth}{\footnotesize{\texttt{gFind(band='both', exponly=True, skypos=[244.27247, 55.268069])}}}}\\\hline
	gAperture & Generates a light curve (returned as a table of times, calibrated fluxes, and additional parameters) for a given coordinate, time sampling, and aperture size. \edit1{Example to create a light curve with 10-second bins in the NUV, with specified aperture radius and background annuli (always in degrees): \parbox[t]{0.315\textwidth}{\footnotesize{\texttt{gAperture(band='NUV', skypos=[244.27247, 55.268069], stepsz=10., csvfile='nuv.csv', radius=0.0045, annulus=[0.0050,0.0060])}}}}\\\hline
	gMap & Creates an image and/or image cubes (in units of counts and/or calibrated fluxes) for a given area of the sky and (optionally) time sampling. \edit1{Example to create an NUV coadd image (in counts) using all available data, as well as an image cube using 30-second slices, of a 6x6 arcmin area on the sky centered on CR Draconis: \parbox[t]{0.315\textwidth}{\footnotesize{\texttt{gMap(band='NUV', skypos=[244.27247, 55.268069], stepsz=30., skyrange=[0.1,0.1], cntfile='imgcube.fits', cntcoaddfile='coadd.fits')}}}}\\\hline
\end{tabularx}
\label{moduledesc}
\end{table}

While the tools have individual syntaxes to fit their specific functions, a few conventions are standard across all of them. Sky positions are reported as two-element vectors (right ascension and declination) in J2000 decimal degrees. Time ranges (or ``bins'') are defined as two-element vectors where the first element is the start time and the second element is the end time, and sequences of time ranges are defined as arrays of such vectors. The gPhoton project defines timestamps in units of ``GALEX time'' throughout, equivalent to a linear offset from UNIX or POSIX time, where $t_{\rm{GALEX}} = t_{\rm{UNIX}} - 315964800$ seconds. To avoid double counting of boundaries, both spatial and temporal ranges are generally taken to be inclusive of the lower value and exclusive of the higher value.

By default, the database tools define an ``effective FoV'' that is 1.1 degrees in diameter, as compared to the full, \edit1{``physical FoV''} of the detector at 1.25 degrees. The effective FoV serves as a means to conservatively trim data that lie near the edges of the GALEX MCPs; \edit1{these regions} suffer from uncorrected, transient edge artifacts and poorly understood sensitivity and spatial distortion. The choice of this effective FoV reflects our suggestion that most users simply avoid data collected near the detector edges. Such data \emph{may} be useful, however, to cautious and knowledgeable investigators, so the effective FoV is adjustable from the command line. Critically, the effective FoV (whether using the default or a custom size) does not eliminate problems caused by photometric apertures, annuli, or requested gMap images that extend into (i.e. are clipped by) the boundary of the \emph{effective} FoV. For reliable photometry, the photometric apertures must not overlap either the physical or effective FoV boundaries. A flag in the gAperture output will alert users to this condition, and it is often visible as a void in gMap movies of the targeted region.

\subsection{gPipeline}
The gPipeline calibration implements a subset of the steps from the original mission pipeline in order to perform detector-level calibration and aspect correction of photon events. The module accepts the raw scientific data file (-raw6), the spacecraft state file (-scst), and one or more refined aspect solution files (-asprta). It returns a ``photon list file'' in CSV format, where each row corresponds to a detector event and records information such as the raw and calibrated detector event positions, sky projected (de-dithered) event positions, and a flag that encodes metadata on the photon event, propagating flags from the aspect solution files and also encoding whether the event falls in a known detector hotspot region. Please see the project documentation for a description of these columns. A flag value of zero at this stage indicates that there were no problems with the calibration of an individual event. The photon list files produced by gPipeline are analogous (but not identical) to the extended photon list (-x) files that were occasionally produced (but not archived) by the mission.

Note that all the inputs to the stand-alone calibration pipeline (-raw6, -scst, and -asprta files) are products from the original mission pipeline that are archived at MAST. By using these archived mission products directly, gPhoton avoids the need to recreate either the ingest or aspect correction stages of the mission pipeline. The relevant content of the aspect and spacecraft state files are also stored in publicly accessible database tables at MAST, allowing gPipeline to be run in either an ``offline'' mode where these files are stored locally or an "online" mode where the software performs web queries to obtain some of the necessary information.

\subsection{gFind}
\label{gfind}
The gFind module allows the user to query the available GALEX exposure of a particular part of the sky. Given a sky position, gFind returns the estimated (raw) exposure depth of available data over the whole mission, separated into time ranges corresponding roughly to discrete observations of the target. Rather than using the visit-based bookkeeping of the mission, which distinguished between observation modes and survey type, gFind uses the photon events themselves. A given position on the sky is considered to be observed if valid data exist in a time range where the position falls within one effective FoV radius of the spacecraft boresight, as defined by the mission-provided aspect solution. Distinct time ranges are identified based on user-adjustable parameters that define the maximum allowable gap between two events for those data to be considered contiguous (or, in other words, part of the same observation) and the minimum raw exposure depth required for an observation to be considered valid.

\subsection{gAperture}
This module extracts and calibrates event-level data from the database to produce light curves, given user-specified parameters that can include target position, photometric aperture, background annulus size, desired integration depth (i.e., bin size), and time range or ranges. Rather than performing photometric measurements on pixelized and integrated images, as the mission pipeline did, gAperture performs aperture photometry by means of cone searches on the sky positions of individual photon events at the native spatial resolution of the data. \edit1{On the client side, each photon event is weighted by the effective exposure time for the whole detector over that time range (Section \ref{effexptime}) and detector flat value at the spot on the detector on which it occurred (Section \ref{relresponsecorr}).} Output from gAperture, which \edit1{includes} a very large number of parameters related to the photometric reduction, can be written to CSV-format tables for later analysis. Of note are columns corresponding to \edit1{time} bin ranges, effective exposure time, intermediate values such as total number of events within the aperture, calibrated source brightness in counts, physical flux, and AB magnitude units derived with a number of background estimation methods (Section \ref{bgcorr}), measurement error (Section \ref{fluxuncert}), and warning flags for a number of conditions that may bias photometric results. Please see the project documentation for a description of gAperture light curve columns. \edit1{The photon event data---including time stamps, detector and sky positions, and response corrections---can be optionally written to a CSV file for detailed analysis, and are also included in the returned data structure when gAperture is called as a Python module.}

\subsection{gMap}
This module creates integrated images or image sequences (i.e. movie cubes) for targeted regions of the sky and specific time ranges, up to and including full depth coadds. Users can request either ``count'' images, which have not been corrected for exposure time or response (often useful for astrometry, diagnostics, or quick-looks), or ``intensity'' images, which are fully calibrated and suitable for photometric analysis. The images produced by gMap are analogous to the imaging data products produced by the mission pipeline, but with additional flexibility provided by means of user-adjustable parameters (e.g. dimensions, exposure depth and binning, edge trimming). If given a sequence of time ranges or a bin size, gMap will also produce ``count'' and/or ``intensity'' image cubes (i.e. movies), which the original mission pipeline could not produce at full spatial resolution. All images are written in the Flexible Image Transport System \citep[FITS,][]{pen2010} format that include headers populated using the World Coordinate System \citep[WCS,][]{gre2002,cal2002} standard.  As with relative response correction in gAperture, rather than generating a relative response map, the individual events are simply weighted by the flat value assigned to the detector regions on which they fell. In the current release, the exposure depth at the center of field is applied evenly across the whole image. This is not a good approximation in a large number of cases, particularly when the diameter of the image is not a small fraction of the diameter of the detector FoV; a spatially aware exposure time correction is planned for future implementation.

\section{Calibration Tests}
\label{calibration}
We performed both relative and absolute tests of gAperture performance, comparing gPhoton output to that of the mission's merged catalog (MCAT) and standard white dwarf calibration star, LDS749B. In all cases of relative comparisons against the MCAT, the catalog source center positions and observation time ranges were used as inputs to gAperture on a per-visit basis.

For tests of relative astrometry and photometry, a circular photometric aperture with a radius of $6''$ was used, equivalent to the MCAT \emph{APER4} column, and the gAperture background annulus was defined to extend from $30''$ to $90''$. When appropriate, measured source magnitudes (from both gAperture and MCAT) were aperture corrected using the table defined in Figure 4 of \citet{mor2007}; \edit1{the corrections for this aperture size are 0.15 AB magnitude in FUV and 0.23 AB magnitude in NUV. To generate a fair sampling of sources across the mission, ``random''  sky positions (right ascension and declination pairs) were determined by a random number generator to serve} as the centers of $0.1$ degree cone searches of the MCAT for all sources between 14 and 22.5 AB magnitude with less than 5000 seconds of total raw exposure coverage (to avoid biasing the analysis with a small number of sources from a handful of very deep fields). \edit1{gAperture was used to generate photometry for 10,000 sources selected in this way in each band. Any integrations for which gAperture returned a non-zero flag value were excluded.}

In figures that include Gaussian Kernel Density Estimates (KDE), bandwidths were chosen by brute force cross-validation, and are represented by blue curves. The peak \edit1{(``average'')} of the KDE is reported as the peak of the distribution of data. To give a sense of the skew of the distribution, we also report the median value along the same \edit1{axes}. For ease of interpretation, it will be useful to note that a difference of magnitudes can be interpreted as a percent difference in flux under a linear approximation near zero.

The source data for all plots, the commands and scripts used to generate the source data, and the scripts used to create the graphics and results are included as supplements to this paper. Similar resources reside in our Github project repository. We encourage researchers to use these as starting points to generate error analyses appropriate to their specific projects.

\subsection{Relative Astrometry}
\begin{figure}[h!]
\includegraphics[width=0.46\textwidth,keepaspectratio]{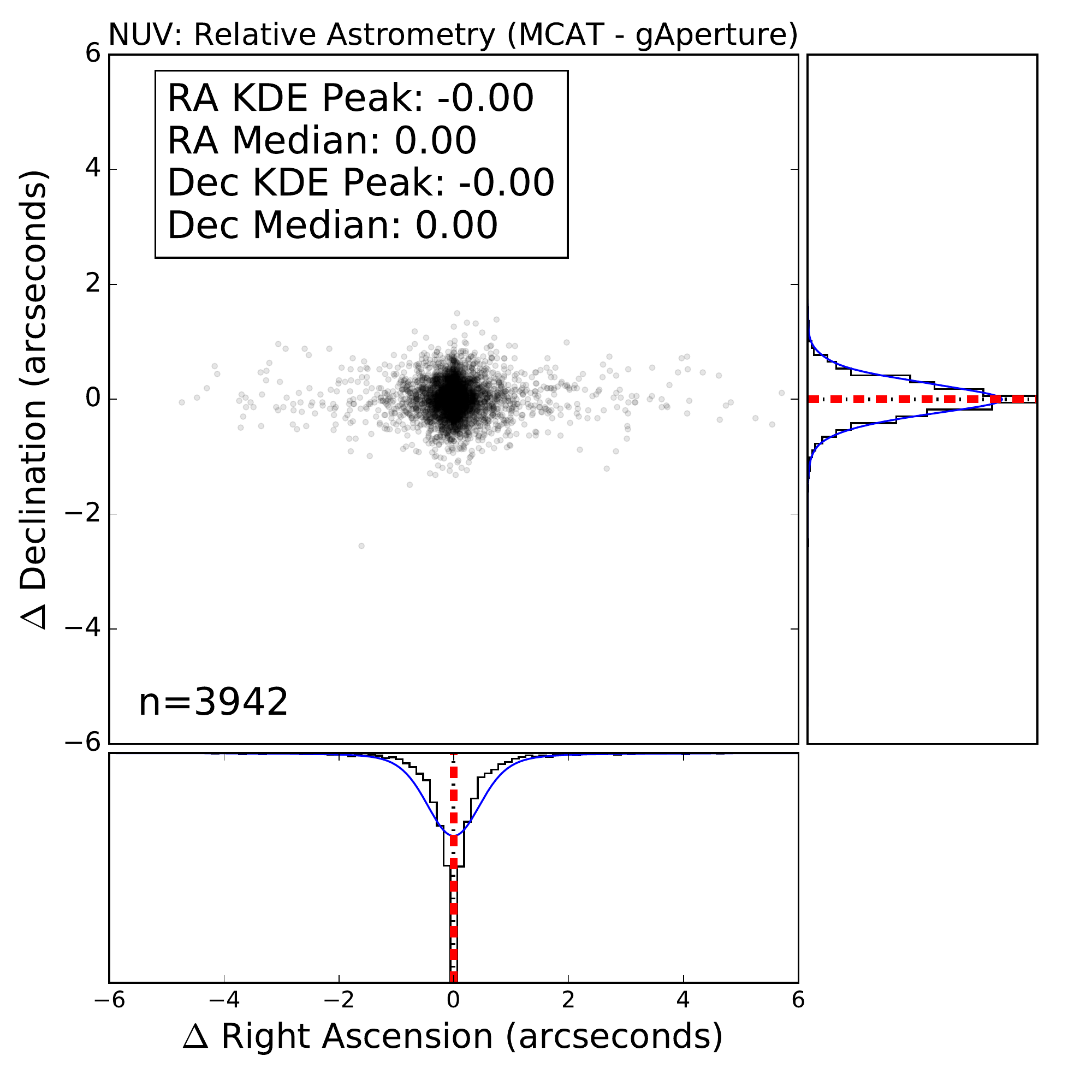}\\
\includegraphics[width=0.46\textwidth,keepaspectratio]{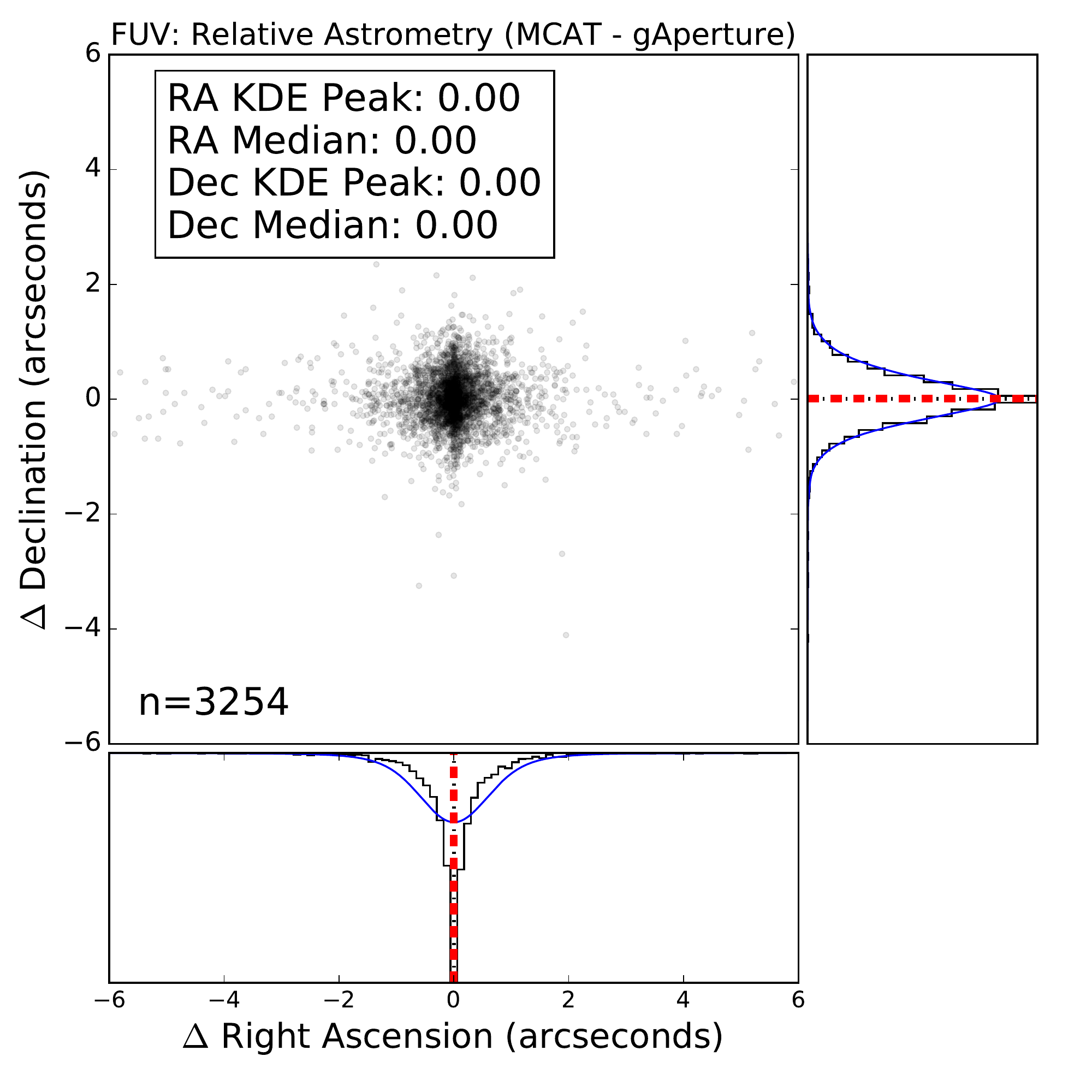}
\caption{Relative offsets between mission catalog (MCAT) source positions and gAperture centers-of-brightness, in the NUV \edit1{(top)} and FUV \edit1{(bottom)}, within photometric apertures with $6''$ radii. \edit2{A Gaussian KDE is overplotted in blue on the histograms of offsets in right ascension in declination. \edit1{On average, t}he relative astrometry is better than $0.01''$ in both bands.}
\label{astrometry}}
\end{figure}

In Figure \ref{astrometry}, we compare the MCAT source center positions to the centers-of-brightness (that is to say, the mean photon position) within a $6''$ aperture. The relative astrometry is very good, with sharp and symmetrical distributions around zero in both \edit2{right ascension and declination} for both bands. A possible cause of divergence in astrometry between gAperture and the MCAT is that the center of brightness was calculated by the mission pipeline on an image with $1.5''$ pixels, necessarily requiring interpolation, whereas gAperture directly samples the detector positions of the incident photons.

\subsection{Background Correction}
\label{bgcorr}
At present, gAperture implements two methods to estimate sky background. The first method uses the background values reported in the \emph{NUV\_skybg} and \emph{FUV\_skybg} columns of the mission-produced MCAT catalog on a per-visit basis. That is, gAperture searches the MCAT for the nearest source to the targeted sky position at the requested time. The mission-produced background flux per area recorded for this source is scaled to the aperture and subtracted from the source flux measured by gAperture. Very broadly, the background estimation procedure in the mission pipeline\footnote{\url{http://www.galex.caltech.edu/DATA/gr1_docs/Background_determination_and_source_extraction_for_GALEX_data.pdf}} used an iterative ``sigma-clipping'' method modified to make probability cuts based on the full Poisson distribution when count rates are low. \edit1{In rare cases that no corresponding visit photometry can be found, \texttt{NaN} is used for this particular output column.}

The second background method implemented by gAperture is an annulus estimate where the surface flux within a user-defined annulus surrounding the extraction aperture is scaled to the area of the aperture and subtracted from the source flux. The annulus background method can produce biased results in cases where it captures light from relatively bright nearby sources, although that effect can \edit1{sometimes} be mitigated by carefully defining the annulus to avoid \edit1{known sources. We suggest that researchers routinely check the MCAT as well as a gMap-produced images of the targeted regions for nearby sources that might bias gAperture photometry.} As discussed further in Section \ref{deadtimedesc}, the diffuse sky background in GALEX observations can vary over the course of an eclipse due to changes in the ambient terrestrial airglow as the spacecraft traveled from limb to limb. When constructing light curves with the intention of looking for short time domain variability, the annulus background method will correct for this variable background, \edit1{whereas the MCAT method cannot.}

\begin{figure}[h!]
\includegraphics[width=0.46\textwidth,keepaspectratio]{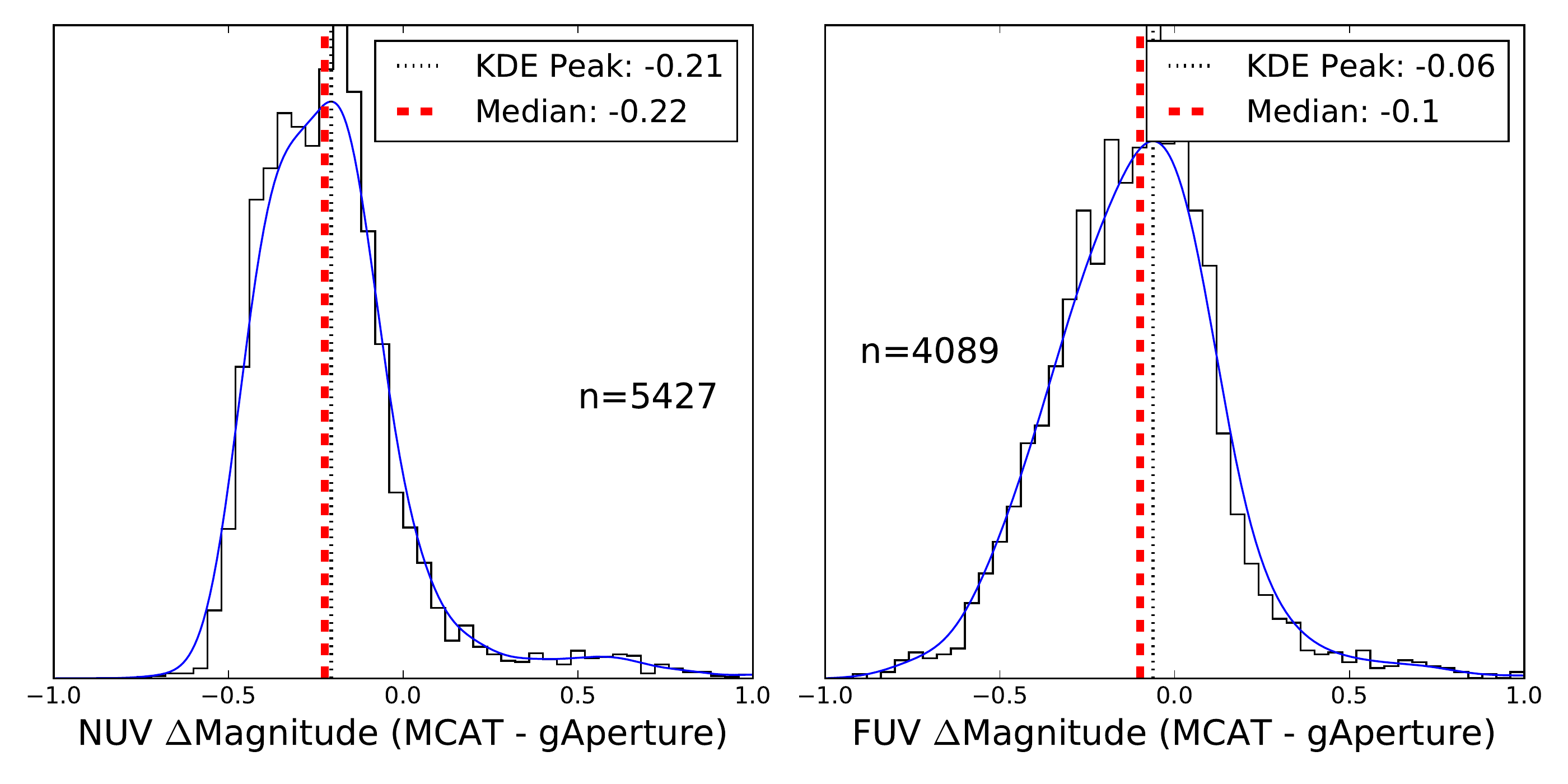}
\caption{\edit2{Histograms of effective background surface brightness within a $6''$ aperture for the gAperture unmasked annulus method as compared to values in the visit-level MCAT with Gaussian KDEs overplotted in blue.} Estimates are within \edit1{$\sim 21\%$} on average in NUV (left) and \edit1{$\sim 6\%$} on average in FUV (right). \edit1{The long tail in each distribution is attributable to contamination of the annulus by relatively bright background sources.}
\label{bgrelphot}}
\end{figure}

A comparison of effective magnitude of the estimated background within a $6''$ aperture as produced by the two methods is presented in Figure \ref{bgrelphot}. \edit1{The Annulus background method produces background estimates that are consistently dimmer than the MCAT, by $\sim 10\%$ in FUV and $\sim 22\%$ in NUV. A disparity between the methods is not unexpected given that the methods are algorithmically quite distinct, but we do not know the specific causes of this difference. A difference in the severity of the disparity between the two bands is also not unexpected given that the NUV band generally has both higher source density and background (\citep{bianchi2011galex}). The offset is quite small, though, in absolute terms: at typical GALEX sky background levels, it amounts to $< 0.03$ counts per second (cps) in NUV and $< 0.02$ cps in FUV.}

\edit1{The long tails in both bands at large magnitude differences are dominated by observations where background stars contaminate the annuli.} During development, we explored two additional methods that might have mitigated the presence of stars in or near the background annulus. The first, which we called ``swiss cheese,'' mimicked the method used in the analysis of the GALEX standard star LDS749B described in \citep{mor2007}: events corresponding to nearby bright stars (as defined by the MCAT) were masked and excluded from subsequent calculations. The second method was an attempt at a direct port of the ``sigma-clipping'' algorithm used by the mission pipeline. Both of these methods were abandoned because they were computationally complex, sensitive to somewhat arbitrary input parameters, and produced poor agreement with catalog fluxes. \edit1{Further exploration of GALEX background estimation methods, possibly including a revisit of these abandoned techniques, is reserved for future work.}

\subsection{Relative Flux Precision}
\label{relflux}
\begin{figure}[h!]
\includegraphics[width=0.46\textwidth,keepaspectratio]{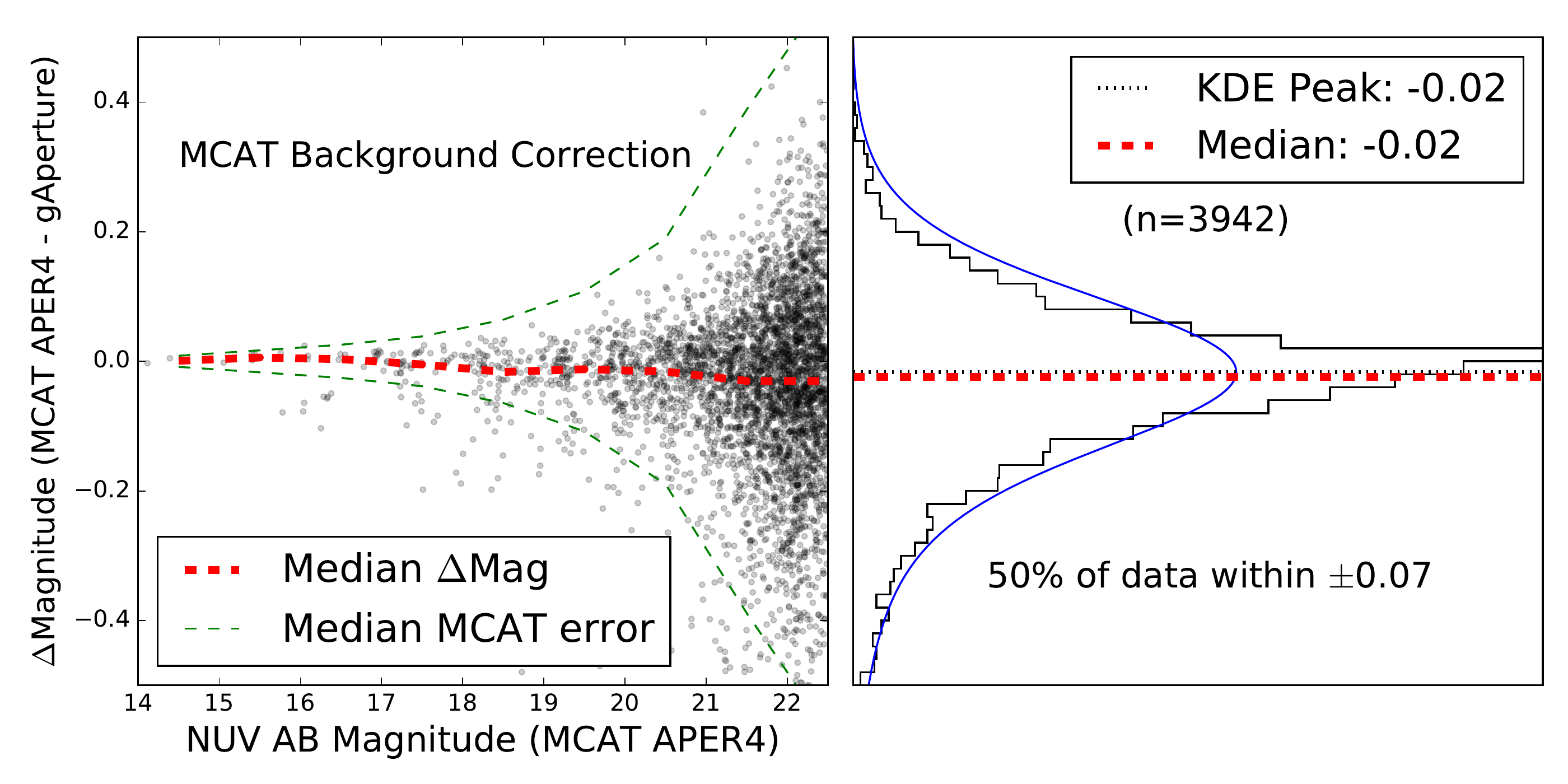}\\
\includegraphics[width=0.46\textwidth,keepaspectratio]{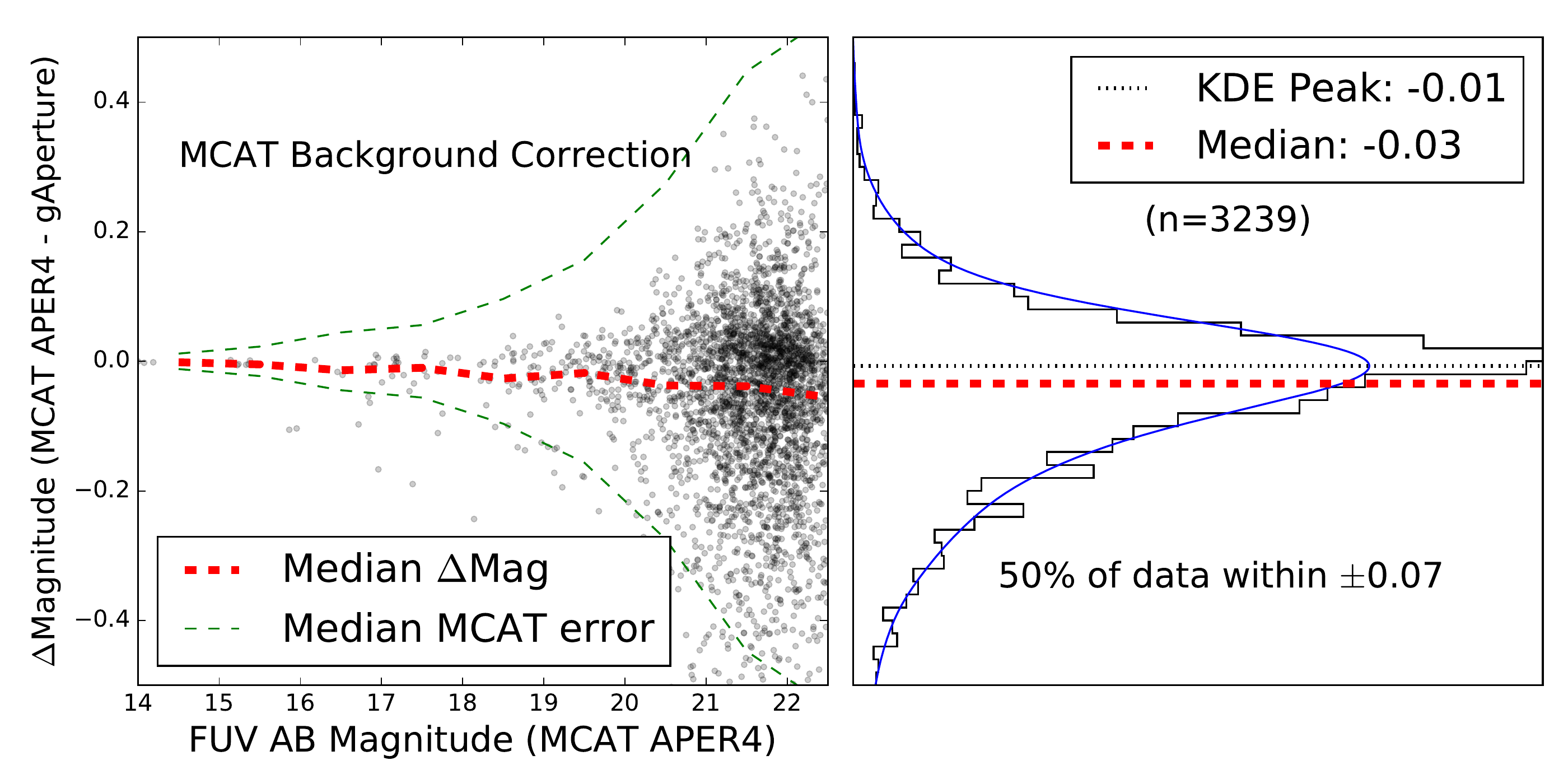}
\caption{Comparison between MCAT and gAperture NUV (top) and FUV (bottom) photometric measurements, using $6''$ radius aperture and backgrounds drawn from the MCAT. The \edit1{thick} dashed red line in the left panels denotes the median difference \edit2{between MCAT and gAperture photometry} in one magnitude bins, \edit1{and the thin dashed green lines are \edit2{median 1$\sigma$ MCAT errors of the data in one magnitude bins}.} \edit2{The right panels contain a histogram of the magnitude differences between gPhoton and the mission pipeline with Gaussian KDEs overplotted in blue.} In NUV, photometry using the MCAT background method agrees within $\sim 2$\% on average, down to 22.5 AB magnitude, and half of the data fall within $\sim 7$\%. In FUV, the photometry using the annulus background method agrees within \edit1{$\sim 1$\%}, on average, down to 22.5 AB magnitude, and half of the data fall within $\sim 7$\%.
\label{mcatrelphot}}
\end{figure}

As a test of the relative photometric precision from gPhoton, we plot the difference between the MCAT magnitude and gAperture magnitude against the MCAT magnitude for randomly selected MCAT sources in FUV and NUV. The sample sizes for the two methods is slightly different, even though the analysis is drawn from the same source data, because the annulus background method can sometimes result in \emph{undefined} magnitudes when \edit1{a background estimate that is brighter than the source} results in a negative \edit1{overall} flux estimate. \edit1{Accurate error estimates for the difference between magnitudes using these two methods would be difficult to compute because the data are not independent, so the errors \edit2{delimited in green are the median MCAT errors in one magnitude bins for the \emph{APER4} values of the data used}, which must be a lower bound on the combined error on the difference in magnitudes.} When using the per-visit MCAT background method, FUV agrees within $\sim 4$\% and NUV within $\sim 2$\%, with good symmetry even for dim sources (Figure \ref{mcatrelphot}).

\begin{figure}[h!]
\includegraphics[width=0.46\textwidth,keepaspectratio]{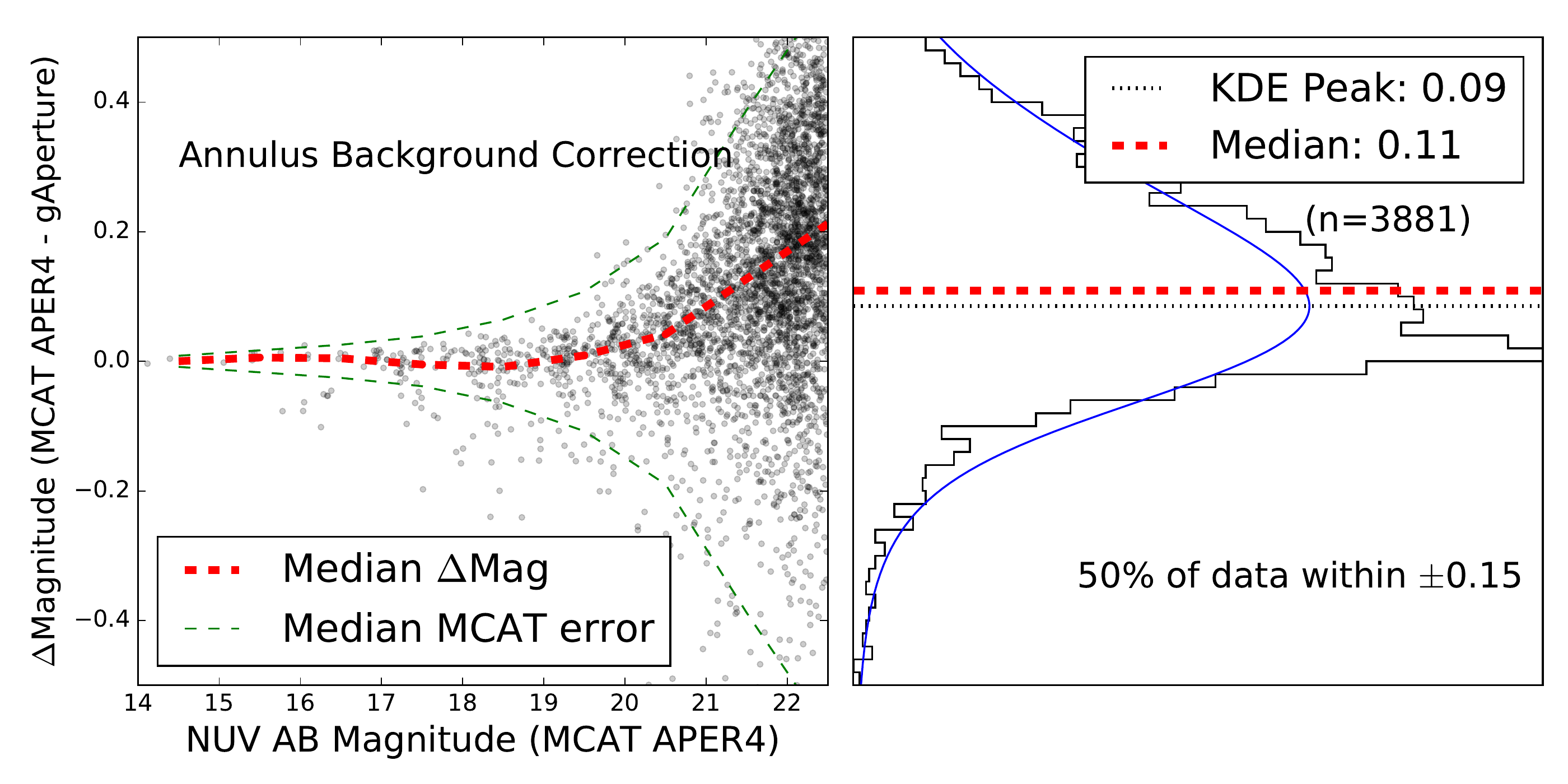}\\
\includegraphics[width=0.46\textwidth,keepaspectratio]{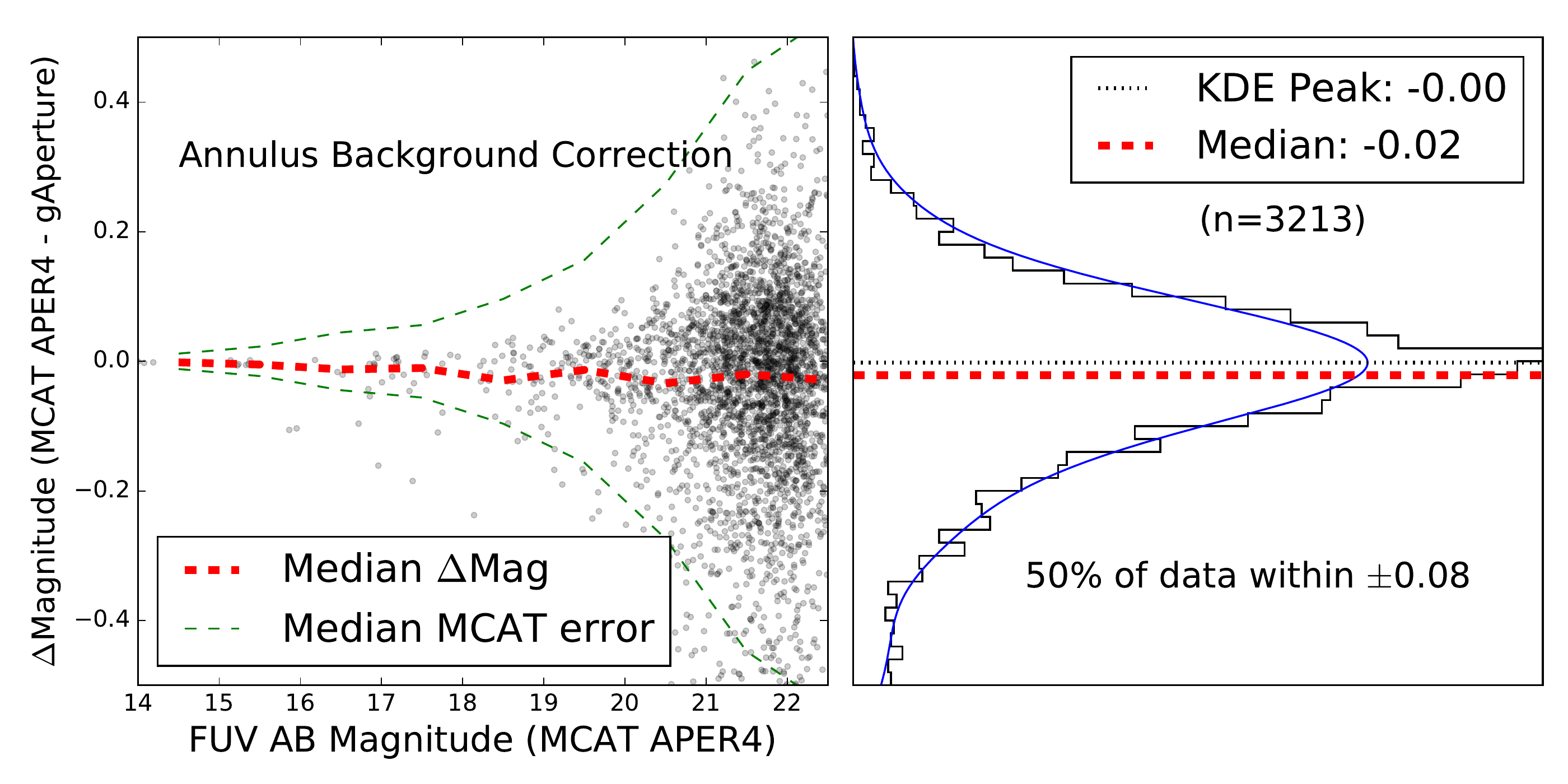}
\caption{Comparison between MCAT and gAperture NUV (top) and FUV (bottom) photometric measurements, using $6''$ radius aperture and backgrounds estimated from unmasked annuli extending from $30''$ to $90''$. The \edit1{thick} dashed red line in the left panels denotes the median difference \edit2{between the MCAT and gAperture photometry} in one magnitude bins, \edit1{and the thin dashed green lines are \edit2{median 1$\sigma$ MCAT errors of the data in one magnitude bins}.} \edit2{The right panels contain a histogram of the magnitude differences between gPhoton and the mission pipeline with Gaussian KDEs overplotted in blue.} In NUV, photometry using the annulus background method \edit1{is consistent with no difference between 14 and 19 AB magnitude}, with a notable increase at dimmer magnitudes. \edit1{NUV photometry agrees with the MCAT to $\sim 9$\% on average, down to 22.5 AB magnitude}, and half of the data fall within $\sim 15$\%. In FUV, the photometry using the annulus background method agrees within \edit1{$1$\%}, on average, down to 22.5 AB magnitude, and half of the data fall within $\sim 8$\%. The divergence for NUV dim sources is due to background, consistent with Figure \ref{bgrelphot}.
\label{annulusrelphot}}
\end{figure}

When using the annulus method, FUV agrees with the MCAT within $\sim 2$\% and NUV within $\sim 11$\% (Figure \ref{annulusrelphot}). The NUV agreement is worse for dimmer sources, consistent with the result in \ref{bgcorr}. \edit1{A similar disparity at high magnitudes does not also show up in FUV because, first, the difference between the two background methods is twice as severe in NUV and, second, typical NUV backgrounds range from $20$-$23$ magnitude (within a $6"$ aperture) with a peak at $\sim 22$ AB magnitude, whereas typical FUV backgrounds range from $21$-$25$, with a peak $\sim 23.5$, which is off the right edges of the left panels of Figure \ref{annulusrelphot}).}

\subsection{Absolute Flux Precision}
\begin{figure}[h!]
\includegraphics[width=0.46\textwidth,keepaspectratio]{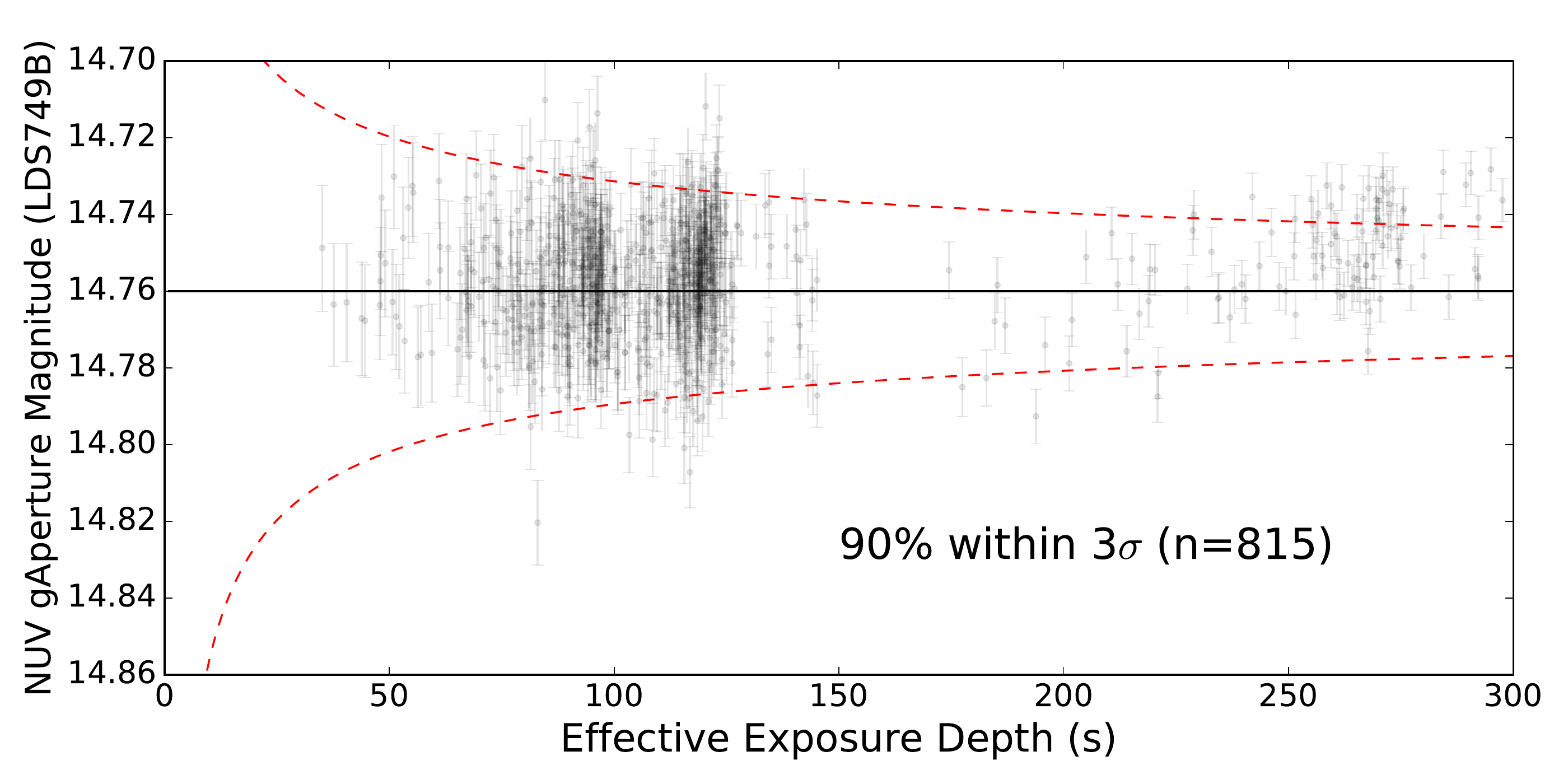}\\
\includegraphics[width=0.46\textwidth,keepaspectratio]{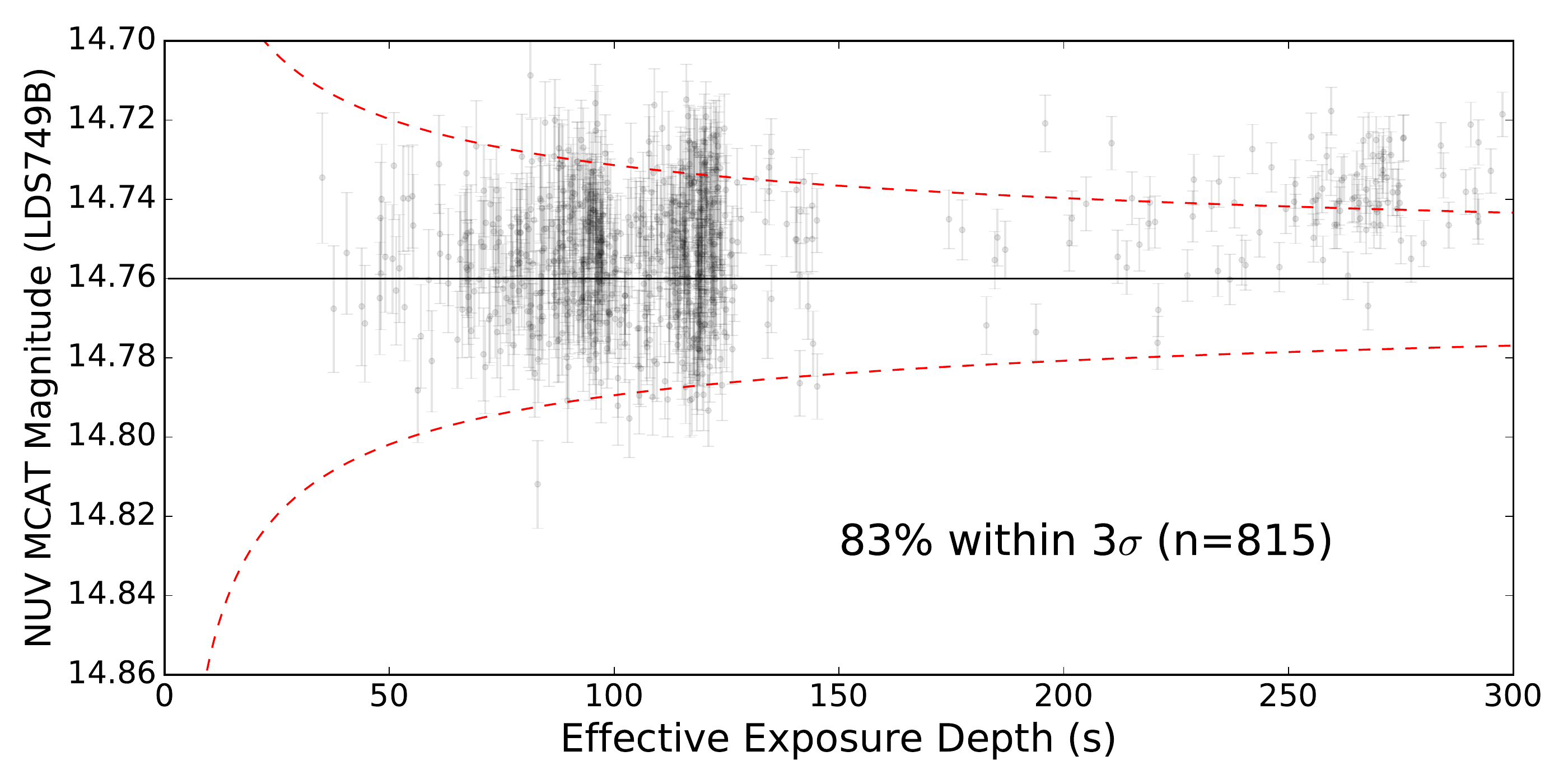}
\caption{NUV photometry of LDS749B as calculated by gAperture (top) and extracted from the MCAT (bottom), using a $17.3''$ radius aperture and MCAT backgrounds with $1\sigma$ error bars. The solid lines correspond to the reference AB magnitude of $14.76$, and the dotted lines denote idealized $3\sigma$ errors as a function of exposure time assuming the reference magnitude and no contribution from background. \edit1{$90\%$} of the gAperture values fall within $3\sigma$ of the reference magnitude, compared to \edit1{$83\%$} of MCAT values.
\label{ldsabsphotnuv}}
\end{figure}

\begin{figure}[h!]
\includegraphics[width=0.46\textwidth,keepaspectratio]{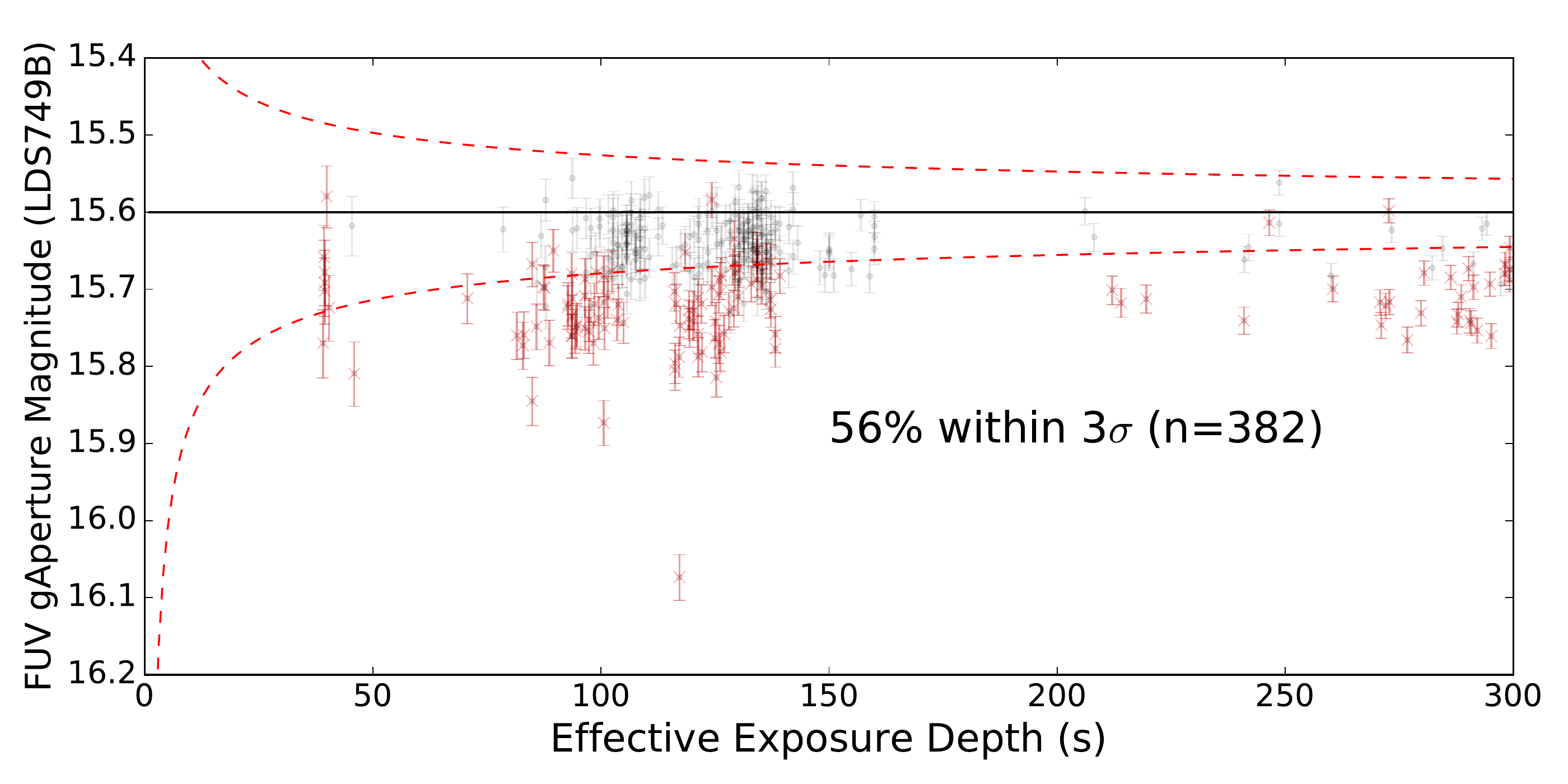}\\
\includegraphics[width=0.46\textwidth,keepaspectratio]{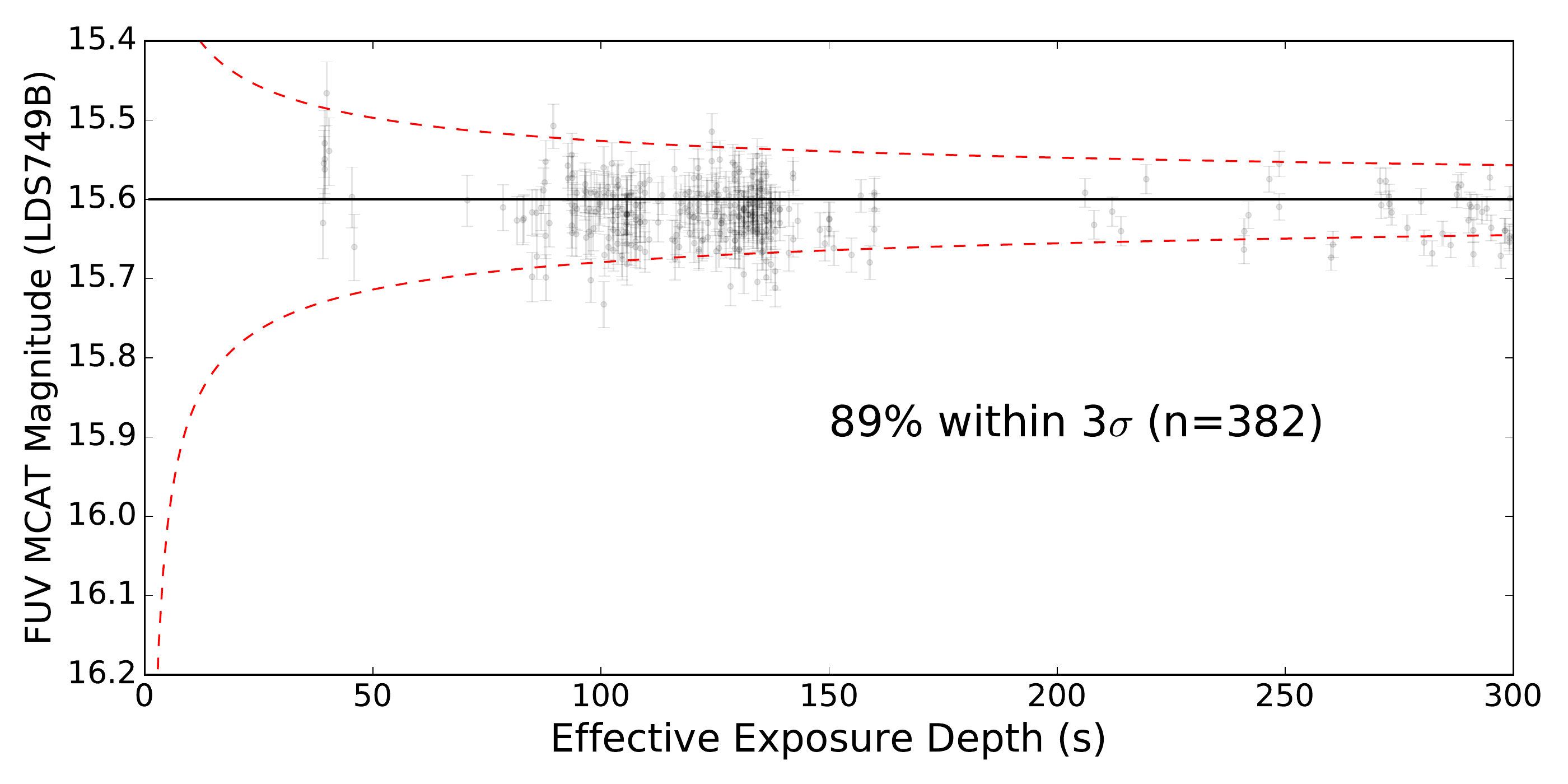}
\caption{FUV photometry of LDS749B as calculated by gAperture (top) and extracted from the MCAT (bottom), using a $17.3''$ radius aperture and MCAT backgrounds with $1\sigma$ error bars. The solid lines corresponds to the reference AB magnitude of $15.60$, and the dotted lines denote idealized $3\sigma$ errors as a function of exposure time assuming the reference magnitude and no contribution from background. \edit1{$56\%$} of the gAperture values fall within $3\sigma$ of the reference magnitude, compared to \edit1{$89\%$} of MCAT values. The points denoted by red crosses correspond to extreme outliers known to correlate strongly with observation leg and for which further analysis is provided in Figure \ref{multimodal}.
\label{ldsabsphotfuv}}
\end{figure}

\begin{figure}[h!]
\includegraphics[width=0.46\textwidth,keepaspectratio]{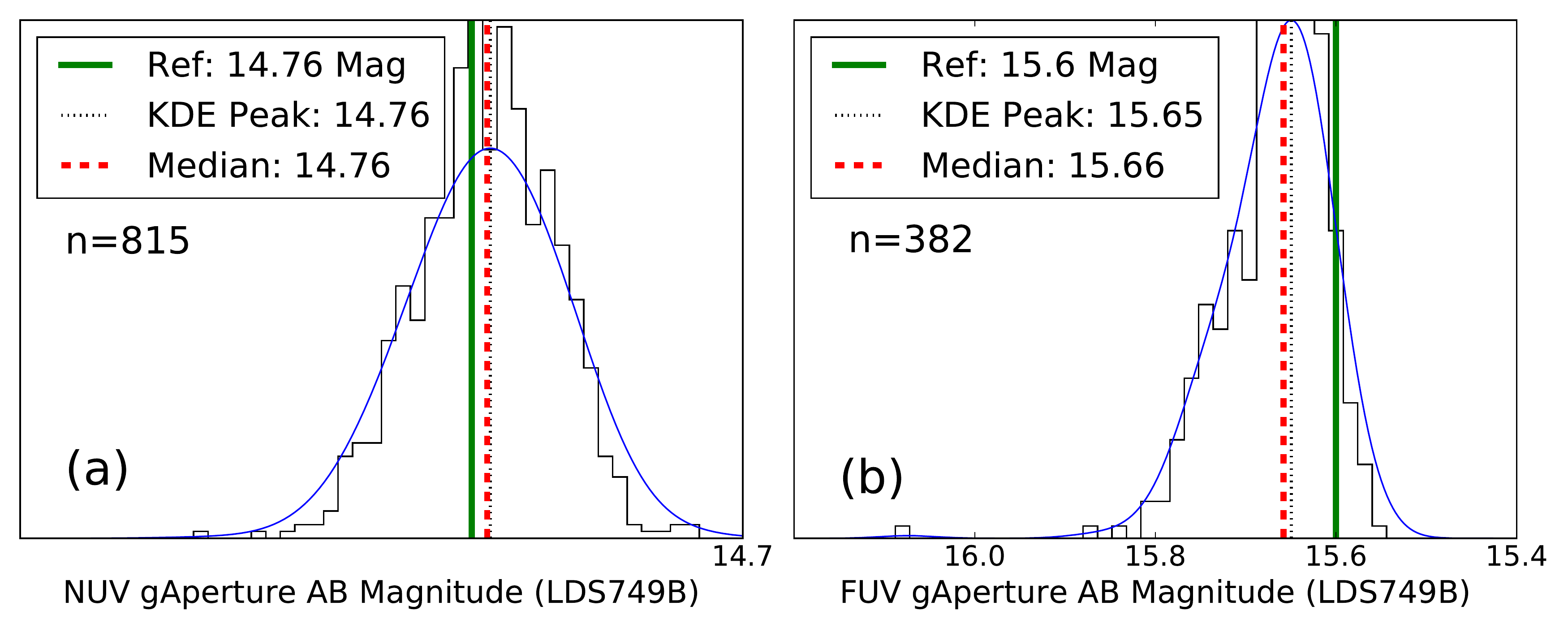}\\
\includegraphics[width=0.46\textwidth,keepaspectratio]{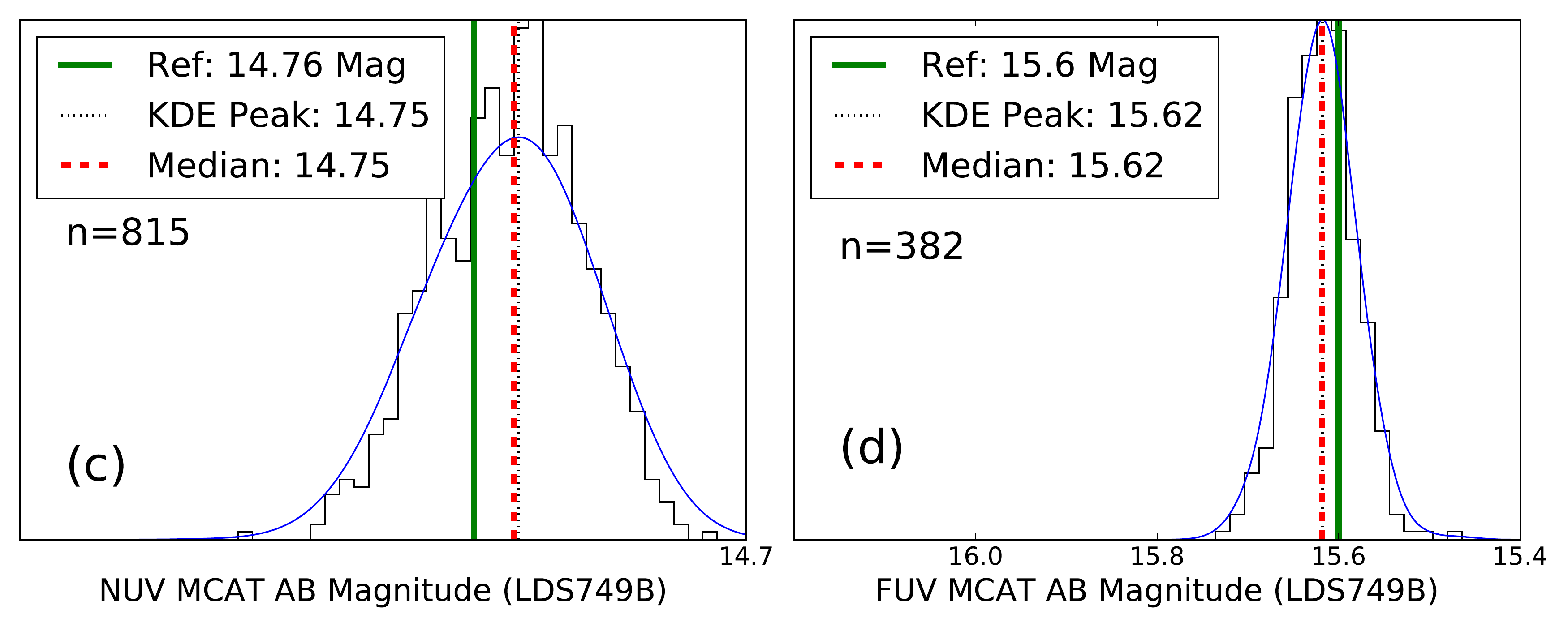}
\caption{Distributions of photometric measurements of LDS749B as calculated by gAperture (top) and extracted from the MCAT (bottom) in both bands, using the same data shown in Figures \ref{ldsabsphotnuv} and \ref{ldsabsphotfuv}\edit2{, with Gaussian KDEs overplotted in blue}. Measurements fall within $\sim1\%$ of the reference magnitude in the NUV for both gAperture (a) and MCAT values (c). Measurements in the FUV using gAperture fall within \edit1{$\sim5\%$} on average (b) compared to \edit1{$\sim2\%$} for the MCAT (d). The larger divergence in FUV is largely attributable to the extreme outliers highlighted in the top panel of Figure \ref{ldsabsphotfuv} and for which further analysis is provided in Figure \ref{multimodal}.
\label{magdist}}
\end{figure}

As described in \citet{mor2007}, the GALEX mission used the white dwarf LDS749B as the primary calibration reference source. We use the refined reference magnitudes of 15.6 AB mag in FUV and 14.76 AB mag in NUV quoted by \citet{camarota2014white} based on the results of \citet{bohlin2008absolute}. The top portions of Figures \ref{ldsabsphotnuv} and \ref{ldsabsphotfuv} display the results of a re-extraction of LDS749B photometry \edit1{by gAperture} as a test of the absolute flux precision. This sample contains all visit-level MCAT detections within 0.001 degrees of the nominal source position, with gAperture parameters set to precisely match time ranges and sky positions in each band. We used a photometric aperture with a $17.3''$ radius, equivalent to MCAT \emph{APER7}, and background estimates from the MCAT. \edit1{The aperture correction in both bands was 0.07 AB magnitude.} To provide a high quality sample, only those sources were considered that did not have a gAperture flag, fell within $1200''$ of the detector center, and were observed prior to the CSP; this cut resulted in a final sample of \edit1{382} visits in FUV and \edit1{815} in NUV. We compare the distributions of fluxes to the predicted 3$\sigma$ counting error, as a function of exposure time, assuming the reference magnitude, \edit1{the aperture correction quoted above,} and no contribution from background. Figure \ref{magdist} provides the magnitude distribution for all observations. \edit1{As a reference,} the bottom panels of Figures \ref{ldsabsphotnuv}, \ref{ldsabsphotfuv}, and \ref{magdist}, contain data for the same observations as pulled directly from the \emph{APER7} column of the MCAT.

In NUV, gAperture produces photometry with a peak density at 14.76 AB mag, with \edit1{90\%} of the data falling within 3$\sigma$ of the reference value of 14.76 AB mag. In comparison, the MCAT values peak at 14.75 AB mag, with \edit1{83\%} falling within 3$\sigma$. In FUV, gAperture produces photometry of LDS749B with a peak density at 15.65 AB magnitude, with \edit1{56\%} of the data falling within 3$\sigma$ of the reference value of 15.6 AB mag. In comparison, the MCAT values have a peak of 15.62 AB mag, and \edit1{89\%} fall within 3$\sigma$ of the reference value.

\begin{figure}[h!]
  \includegraphics[width=0.46\textwidth,keepaspectratio]{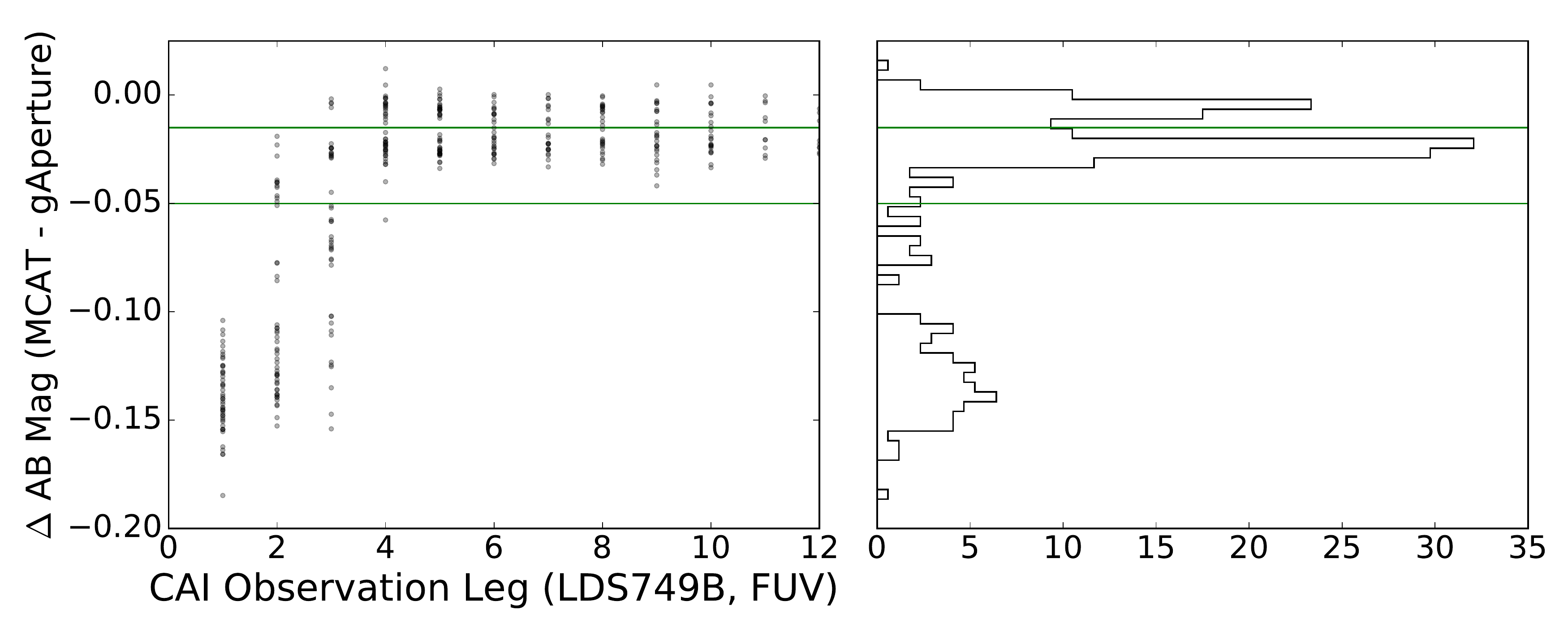}\\
  \includegraphics[width=0.275\textwidth,keepaspectratio]{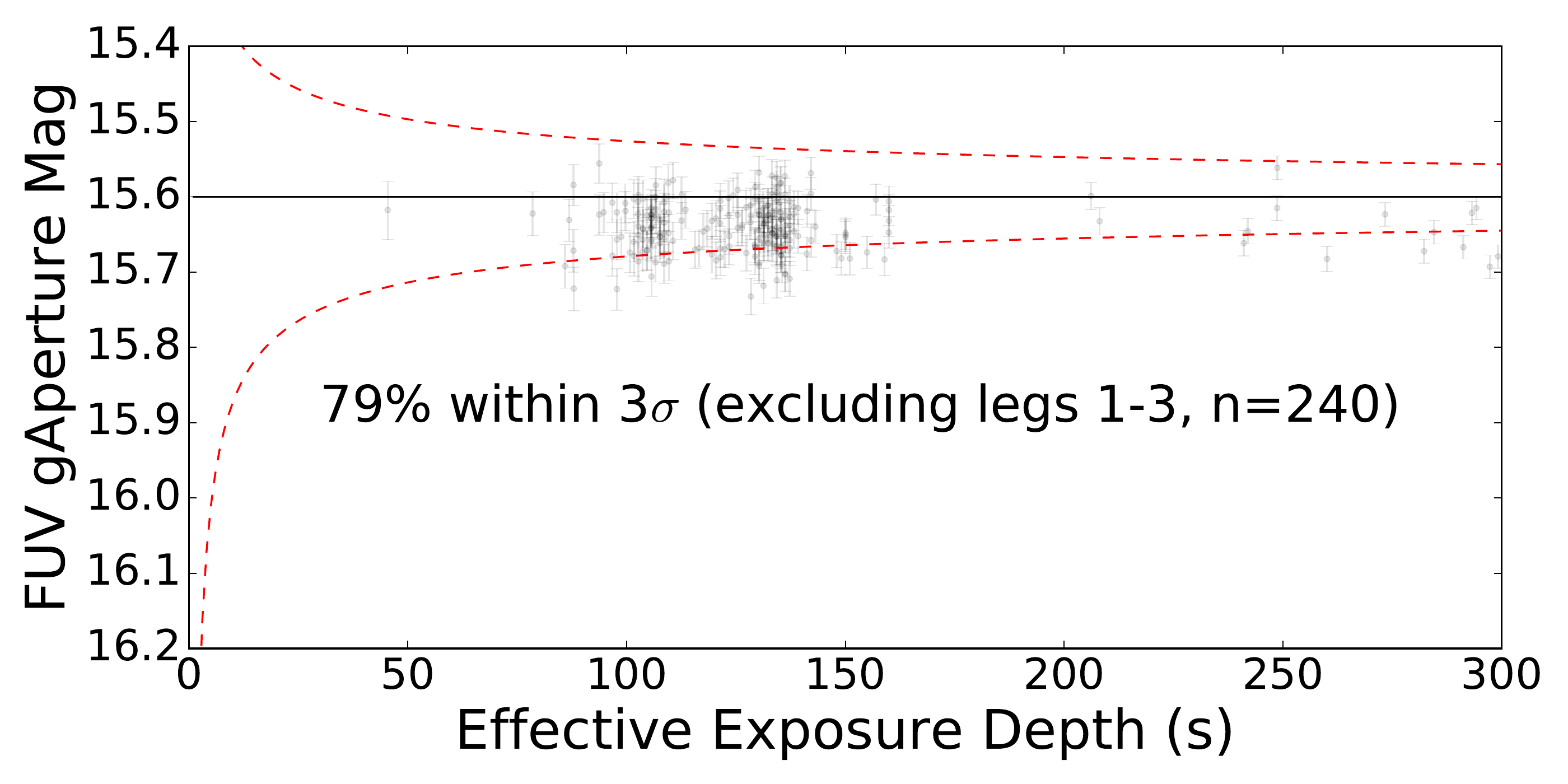}%
  \includegraphics[width=0.172\textwidth,keepaspectratio]{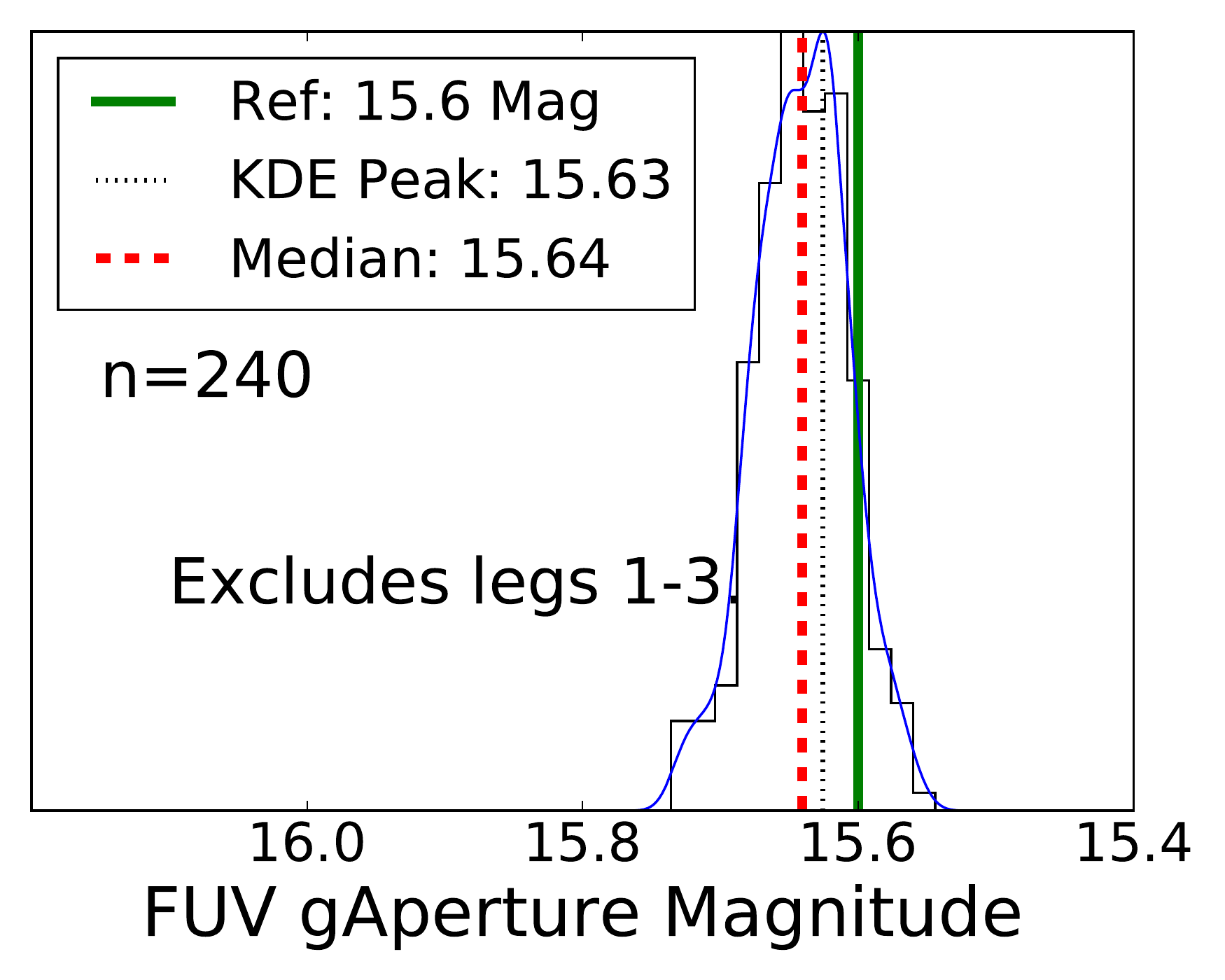}
  \caption{(Top) There is a multi-modal distribution in the difference between gAperture and MCAT magnitudes of some sources, especially obvious for relatively bright stars observed as part of the calibration survey (CAI). Modes appear to be centered around offsets of 1\%, 3\%, and 15\%, with gAperture consistently producing dimmer photometry than the mission. The worst of these outliers are strongly correlated with the first three legs of the calibration aspect pattern, but the cause is currently unknown. (Bottom Left) When observations occurring in the first three legs of the CAI pattern are excluded from analyses, \edit1{79\%} of the remaining \edit1{240} FUV data points lie within 3$\sigma$ of the LDS749B reference value of 15.6 AB magnitude. (Bottom Right) With the first three legs of CAI pattern excluded, the distribution of the remaining measurements of LDS749B fall within \edit1{$\sim3\%$} of the reference value based on the peak of the Gaussian KDE which is overplotted in blue. Note that \edit2{there is still a double peak} in the distribution, corresponding to the two modes with smaller offsets. The larger of these remaining modes corresponds to a bulk shift in photometry between calibration data observed before September of 2007 and after July of 2008.
  \label{multimodal}}
\end{figure}

We have not yet been able to determine the cause of the larger dispersion in FUV (Figure \ref{ldsabsphotfuv}, top). It is easily detected as a multi-modality in magnitude differences between gAperture and MCAT photometry of the same source. The photometry produced by gAperture consistently reports dimmer values than the MCAT, with clusters at offsets of approximately 1\% and 3\% and between 5\% and 15\%. There is a bulk offset of a few percent between calibration data collected before November of 2007 and after June of 2008, which accounts for the modes at 1\% and 3\%. While dates are consistent with recovery from an FUV anomaly in November 2007 and an instrument shutdown in June of 2008, we have no explanation for why these would create a discrepancy between gPhoton and MCAT photometry. The more severe outliers at $\sim10\%$ are strongly correlated with observation leg number. In order to thoroughly sample the detector with the calibration standards, many calibration survey (CAI) observations were collected in petal-pattern mode, but each individual leg was processed as a unique visit in the style of AIS. The most extreme of these outliers are almost exclusively confined to the first three legs of these observation sequences (Figure \ref{multimodal}).

We have found this multi-modal behavior in some but not all other brighter sources in the LDS749B field and in observations of the LB227 calibration standard. We have not found clear evidence of the multi-modality in deep ``dither-style'' observations of other FUV sources, but there are not sufficient observations of a single source in AIS-mode in any survey outside of CAI for us to say with certainty. The fact that no such multi-modality shows up in the tests of relative flux precision (Figure \ref{annulusrelphot}, top) suggests that the problem might be limited to CAI data. Approximately 2.24\% of all FUV observations, by exposure time, are in CAI mode, which is less than 1\% of all GALEX observations. When observations from the first three legs are excluded, gAperture produces FUV photometry of LDS749B with a peak density at 15.63 AB magnitude and \edit1{79\%} of the data falling within 3$\sigma$ of the reference value of 15.6 AB mag. With the most severe outliers filterable by leg (down to about the \edit1{3\%} level) and less severe outliers filterable by time, this issue should not preclude the use of most gAperture FUV photometry. \edit1{Please consult the User Guide for details on how to check whether data fall within these legs.} Understanding and addressing this issue is a top priority for future work.

\section{Implementation Notes}
\label{implementation}

\subsection{Effective Exposure Time}
\label{effexptime}
The GALEX ``effective exposure time'' is defined as the raw exposure time, minus the amount of time considered ``shuttered,'' scaled by the global dead time ratio.

\[t_e=(t_r-t_s)*d\]

The raw exposure time ($t_r$) is computed with the same algorithm used by \emph{gFind} (Section \ref{gfind}). Any time period of 0.05 seconds or longer during which no valid data was recorded by the detector (i.e. with photon event database flags of zero) is considered \emph{shuttered}; the sum of time over such periods during the observation is the shutter correction ($t_s$). These might be periods during which the spacecraft was not actually observing the requested region of sky, but can also include data dropouts or periods during which a valid aspect solution is not available. The global deadtime ratio ($d$)---described more completely in the Section \ref{deadtimedesc}---is the estimated fraction of time during which incident events were \emph{missed} due to detector readout. For aperture photometry, the effective exposure is computed at the targeted sky position, and then applied uniformly across all events in both the aperture and background annulus. This approximation is more efficient than calculating the exposure across the whole region, and fails only when the annulus or background contains a masked part of the detector (e.g. hotspots, as in Section \ref{hotspot}) or crosses the edge of the FoV; \edit1{these conditions are automatically detected and flagged by gAperture.}

\subsection{Exposure Dead Time Correction}
\label{deadtimedesc}
Micro-channel plates are subject to a global exposure ``dead time'' effect caused by the inability of the detector to process more than one event at a time. That is, while a single event is being recorded by the detector electronics, other incident events go undetected. The effect scales inversely as a function of total global detector count rates, or the totality of all events (both null and non-null) recorded by the detector: as global count rate \edit1{increases}, the fraction of exposure lost to dead time likewise increases. For normal GALEX observations, the global count rate is dominated by the observed field brightness, and the relationship between global count rate and dead time is linear.

The GALEX detectors were equipped with four built-in electrical pulsers (``stims'') located off the main detection window that produced a known rate of events, \edit1{nominally 79 cps total between the four}. The mission pipeline estimated a correction by observing that the ratio of the measured stim count rate to the nominal stim count rate should be the same as the ratio of the effective exposure time to the raw exposure time. While this "deadtime ratio" varies quite a bit between and even within observations, a typical value is around 0.8 (indicating that 20\% of exposure time is ``lost'' to dead time).

While the stim rate technique used by the mission works well over long integrations, it introduces unacceptable error in exposure time estimates over \edit1{short} integrations. At the typical deadtime value of 0.8 noted above, the detected stim count rate (across all four stims) would be approximately 63.2 cps (80\% of the nominal rate of 79 cps). For an AIS-depth integration of 100 seconds, over which $\sim 6320$ stim events would be detected, the 1$\sigma$ counting error in the stim measurement would be $\sim 79.5$ counts, corresponding to 1.25\% error in the estimated exposure time. This is small compared to other sources of uncertainty in the imaging chain, and, indeed, was not even propagated by the mission pipeline. However, at the more rapid cadences enabled by the gPhoton architecture, this error becomes significant. For example, for 1 second exposures the 1$\sigma$ error on the stim count rate amounts to 12.6\%.

gPhoton mitigates this by using the linear relationship between global count rate and dead time, which holds for the majority of global count rates observed by GALEX up through GR7, to produce an empirical exposure time correction as a function of global count rate. GALEX has typical global count rates of 10,000 cps or more, making the 1$\sigma$ error due to counting statistics truly negligible even for short integrations. The GALEX team did produce (but did not publish) such an empirical dead time formula; while the result was recorded, the actual methodology was not, making it impossible to verify \citep{mor2013}. We know that the behavior of the two detectors was deemed in that analysis to be sufficiently similar that an identical model fit was used to describe both of them, and the nominal / commanded stim rate was assumed to be \emph{true} (i.e. the stim rate at ``zero'' global counts was fixed to 79 cps). For completeness and consistency, we have redone this empirical deadtime analysis without those two assumptions.

\begin{figure}[h!]
\includegraphics[width=0.46\textwidth,keepaspectratio]{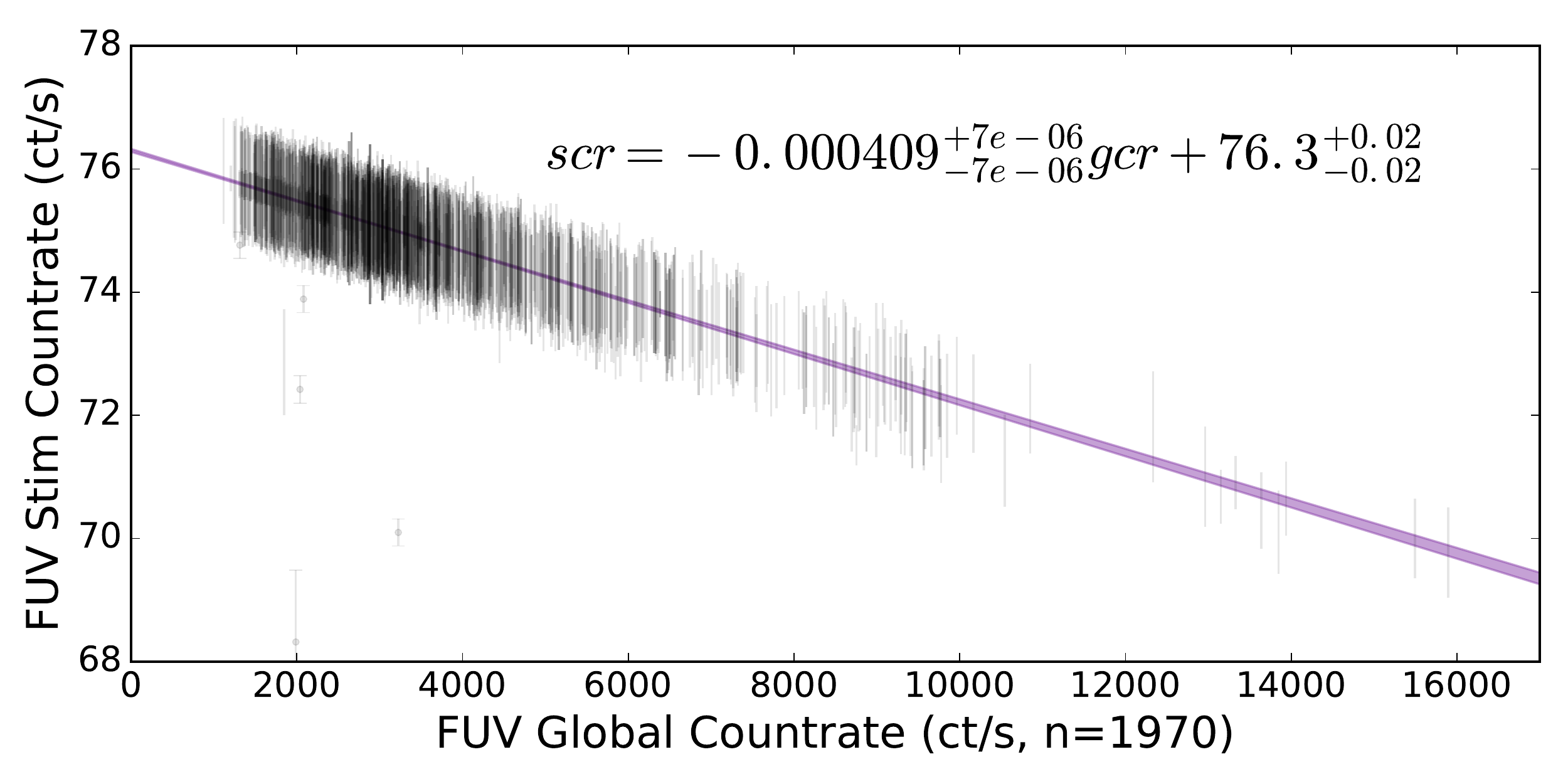}\\
\includegraphics[width=0.46\textwidth,keepaspectratio]{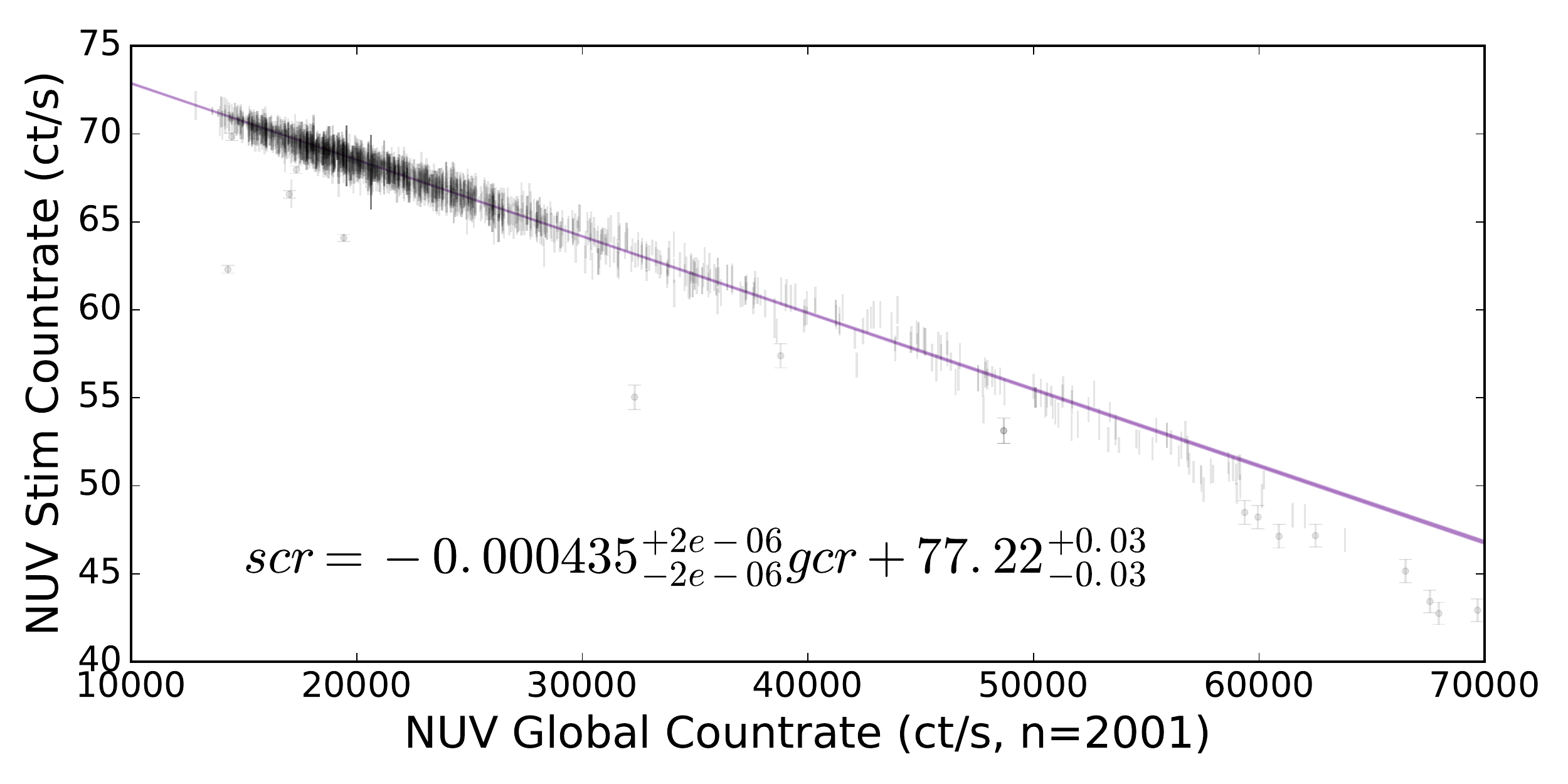}
\caption{FUV (top) and NUV (bottom) stim rates and fit of a linear mixture model to stim count rates \edit1{(scr}) as a function of global count rate \edit1{(gcr)}. Values for $16\%$, $50\%$ and $84\%$ confidence intervals in offset and slope are provided. The very small number of data points indicated by a filled in circles were rejected by the mixture model and not used in the fit. The non-linear rolloff of data starting near 60,000 cps in NUV is likely a real effect due to global detector gain sag.
\label{stim}}
\end{figure}

In Figure \ref{stim}, we plot global detector count rates against stim count rates with 1$\sigma$ errors based on counting statistics for both bands. In calculating these rates, exposure times have been corrected for shutter (see Section \ref{effexptime}), but not dead time. We fit a linear mixture model to the data, with both ``foreground'' and ``noise'' parameters. The model was sampled by Markov Chain Monte Carlo (MCMC) using the ``emcee'' package \citep{for2013} against the data for $\sim 2000$ observations in each band to produce maximum likelihood model parameters for the stim count rate as functions of global count rates, which can be converted to a fractional dead time by comparing the stim count rate against the reference rate. Rather than directly adopting the quoted stim reference rate of 79 cps for both bands, we used the maximum likely y-intercept value corresponding to ``zero'' global counts per second. At 77.2 and 76.3 cps in NUV and FUV respectively, the maximum likely stim count rates differ by 2.3\% and 3.5\% from the commanded rate of 79 cps, and by 1.2\% from each other. The analysis suggests slopes that differ in each band by about 6.5\%. Less than $1\%$ of data in both bands were classified as noise by the model.

The linear empirical deadtime correction will overestimate effective exposure times for very bright fields (where gAperture will produce erroneously dim flux estimates). The deviation from a linear relationship between the global and stim count rates above 50,000 cps in NUV is likely a real effect due to some combination of the onset of the non-linearity in the dead time correction that is expected at high global count rates or other gain-sag effects inherent to the detector (see \citet{mor2007} for a discussion of these effects). The majority of observations through GR7 fall within the linear regime, but the non-linearity in exposure time correction may be a major consideration for data collected late in the mission or during the CAUSE phase when global and local detector brightness limits were relaxed.

Note that the ratio of effective exposure time to raw exposure time is not constant over \emph{any} finite period. The detector FoV always moved in relationship to the sky, resulting in small changes to field brightness and therefore global count rate. At the same time, the spacecraft is traveling through the shadow of the Earth, encountering shifts in ambient brightness due to airglow. To account for this, gAperture recomputes the exposure time and correction factors independently for each time range or bin.

\subsection{Relative Response Correction}
\label{relresponsecorr}
In the mission pipeline, variable sensitivity (``response'') across the detector was corrected by the application of relative response maps (-rrhr). These maps were composed of successive projections of an upsampled detector flat on the sky in one-second increments, weighted for effective exposure time over those increments. The response maps could then be divided out of the integrated count maps over the same time range to produce fully calibrated intensity maps (-int). In developing gPhoton, we discovered that not only did this repeated interpolation unnecessarily degrade the information in the flat, but it was computationally intensive and slow. For this reason, we apply the flat at the detector level by weighting each individual photon event by the value of the pixel in the uninterpolated flat that corresponds to the detector location at which the event was recorded. The exposure time correction is applied independently.

\subsection{Hotspot Masking}
\label{hotspot}
Hotspots are regions of the detector known to produce anomalously high signals that are not correlated with the observed scene, often due to hardware flaws or damage. Regions of the detector flagged as containing hotspots should not be used in routine data analysis. At present, data that fall within regions of the detector covered by the hotspot mask are not aspect corrected and so these regions present as gaps in coverage as a function of detector position. When a source traverses such a region during an observation, it can appear as significant and time-variable dimming in the light curve that can easily be mistaken for real astrophysical phenomena like pulsation or transits. Users should be extremely skeptical of any variability that correlates strongly with the gAperture flag that indicates a nearby masked hotspot region. Many GALEX hotspots are known to be transient, however, such that the masks often block valid observational data; a planned improvement to the pipeline and database will aspect-correct the masked data and apply the mask on the client side in the same manner as the response correction, at discretion of the user, such that overzealously masked but valid data can be recovered.

\edit1{\subsection{Local Non-linearity}
\label{nonlinear}
\citet{mor2007} reports a local non-linearity in detector response (as distinct from the \emph{global} non-linearity described in Section \ref{deadtimedesc}) with a $10\%$ reduction in flux at $109$ cps in FUV and $311$ cps in NUV, which corresponds to AB magnitudes of $13.73$ and $13.85$ respectively. This condition is flagged in gAperture light curves. We have found that sources near and above the local non-linear regime of the detectors frequently present as false short time-domain variables, often exhibiting significant pulsations over single visits that correlate or anti-correlate with detector position \citep{delavega2016}. A similar issue can arise for dimmer sources near the detector edges, proximity to which also triggers a gAperture flag. Possible variable sources should be screened for this.}

\subsection{Flux Uncertainties}
\label{fluxuncert}
The flux uncertainties provided by gAperture are computed by adding the counting errors in the aperture and background annulus in quadrature, scaled to the area of the aperture. If there are relatively bright sources located in either the aperture or background annulus, then this misestimates uncertainty in proportion to the level of contamination. Before relying on the estimated flux uncertainties, users are encouraged to visually check a full depth coadd of the targeted region (as created by gMap) for nearby sources. Future work will include better modeling of the imaging chain as a means to more accurately propagate uncertainties.

\subsection{Optimal Time Bin Sizes}
\label{optbinsize}
\begin{figure}[h!]
\includegraphics[width=0.46\textwidth,keepaspectratio]{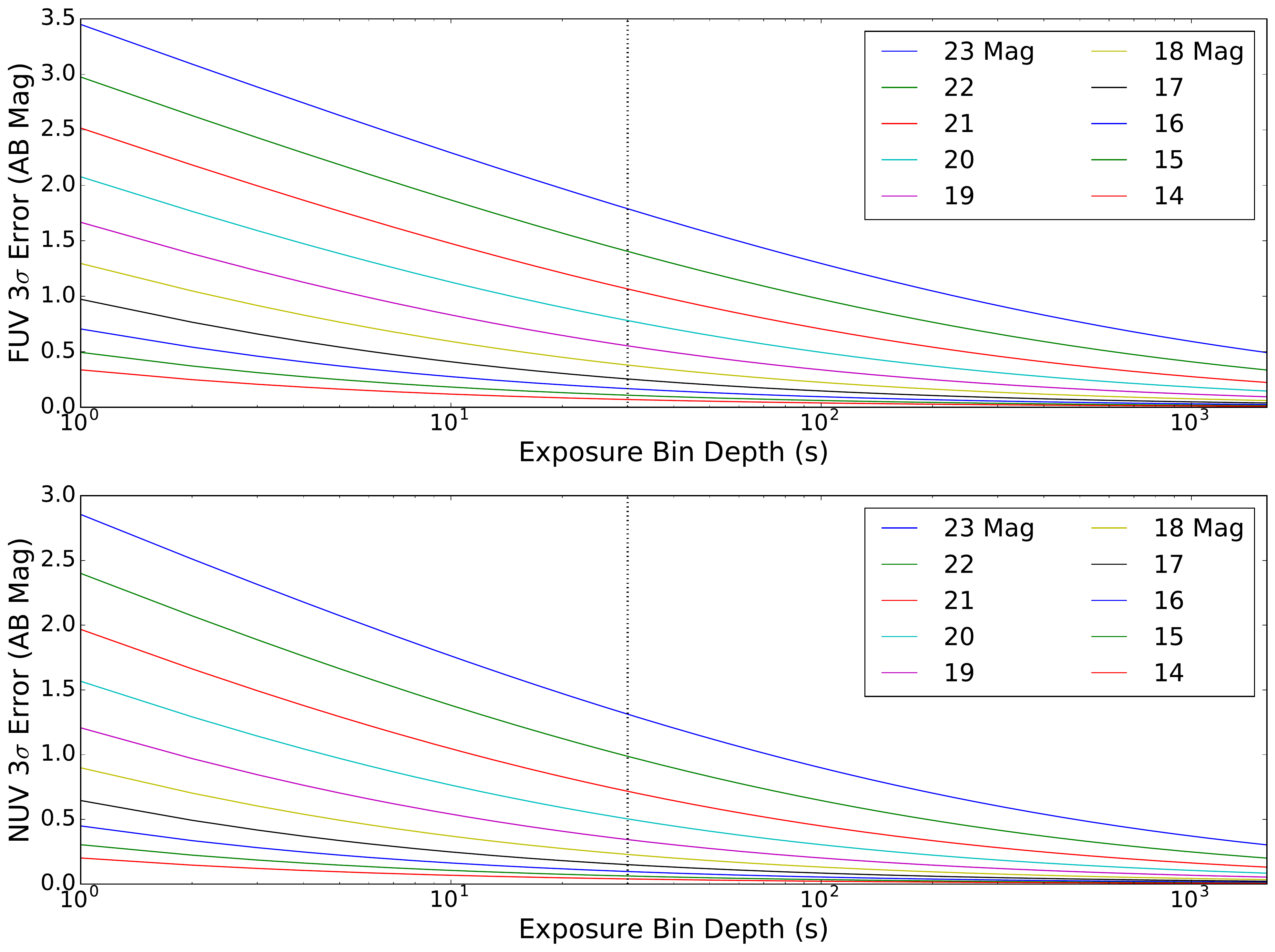}
\caption{Optimal light curve bin sizes depend on a number of factors, most prominent of which are the GALEX band in which the analysis is being performed, the AB magnitude of the target being observed, and the desired precision of the measurement of source brightness or change in source brightness. These figures provide estimates of 3$\sigma$ measurement errors in both bands, based on counting statistics \emph{with no contribution from background}, for a range of target brightnesses and exposure depths up to a full eclipse.
\label{sigmadetlim}}
\end{figure}

The first question of many potential gPhoton users will be whether a temporal phenomenon of interest is actually detectable in the GALEX data using gPhoton. The answer to this question depends on the timescale of the phenomenon in question, the GALEX band of interest, the target brightness, the magnitude of the variability of interest, the local background, the choice of aperture and annulus extents, and the desired measurement uncertainty. Figure \ref{sigmadetlim} presents a model of measurement uncertainty as a function of integration depths for a range of source brightnesses in both bands under an assumption of no background contribution. We recommend that most exploratory analyses begin with a 30 second time bin, as this provides a good midpoint in measurement error between the longest and shortest possible integrations. Any potential variability should be confirmed across several bin depths, to eliminate the possibility of aliasing. The magnitude of potential variable behavior should be carefully assessed in the context of the measurement error and gAperture quality flags.

\subsection{Client vs. Server Optimizations}
\label{speedopt}
In early design concepts, we anticipated that a large amount of data processing would be offloaded to the database and server. In practice, we found it more convenient for both developers and users to conduct the majority of processing on the \emph{client} and reserve server-side operations to standard, straightforward SQL operations like merging tables or counting rows. We were also surprised to discover that the total runtime was frequently not dominated by processing on either the client or server, but by the handling of the \emph{http} requests between the two---that is, the time required just to send and receive a response from the server was a larger factor than the time required to \emph{compute} the result. Significant development work has gone into minimizing the total number of such requests. Among the most substantial and surprisingly effective strategies has been to download almost \emph{all relevant data}---anything within the targeted sky regions and time ranges, which can easily be millions of database rows---to the client early in each run, and performing most subsequent analysis on those data locally.

\section{Example Science Application - Stellar Flares From CR Draconis}
\label{scienceexamples}
CR Draconis (HIP 79796) is a fairly bright ($V \sim 10$) binary star system composed of two M dwarfs located $\sim 20$ pc away in a slightly eccentric orbit with a period of $\sim 4$ years \citep{tam2008}, and has been known to exhibit flares for many decades \citep{cri1970}. The system was \edit1{categorized} as a high-amplitude variable in the second version of the GALEX Ultraviolet Variability (GUVV-2) Catalog \citep{whe2008}, where a maximum NUV flux difference of two magnitudes was identified within the available visits at that time. \citet{wel2006} studied one of CR Draconis' flare events with high temporal sampling by extracting light curves from sky-projected, ``extended'' (-x) photon list files, produced as non-standard products of the GALEX mission pipeline.

Using our gPhoton pipeline, we have searched for flares from the CR Dra system \edit1{in} GALEX data spanning the lifetime of the mission. \edit1{The example calls to the gPhoton methods shown here assume that the modules have been imported to Python by means of \texttt{import gPhoton}. Our first step is to search the entire database to determine how much data is available using \texttt{gFind}:}

\begin{verbatim}
data = gPhoton.gFind(band='NUV',
                     skypos=[244.27246917,
                             55.26919386])
\end{verbatim}

\edit1{Here we search in the NUV band, and do not specify a \texttt{detrad} value, thereby using the default effective field-of-view to avoid any observations where CR Draconis was too near the edge of the detector.  If we wanted to include these, we would specify \texttt{detrad=1.25}, although in general we advise against this.}

\edit1{We next create a coadd image using all the available data using \texttt{gMap}, which we use to search for any faint objects nearby, as well as an image cube of 10-second frames, which we can use to visually check for large-scale variations, identify any image artifacts, and define our photometric apertures.  The command for this is:}

\begin{verbatim}
gPhoton.gMap(band='NUV',
             skypos=[244.27246917,
                     55.26919386],
             stepsz=10.,
             skyrange=[0.1,0.1],
             cntfile='cube.fits',
             cntcoaddfile='coadd.fits')
\end{verbatim}

\begin{figure}[h!]
\includegraphics[width=0.46\textwidth,keepaspectratio]{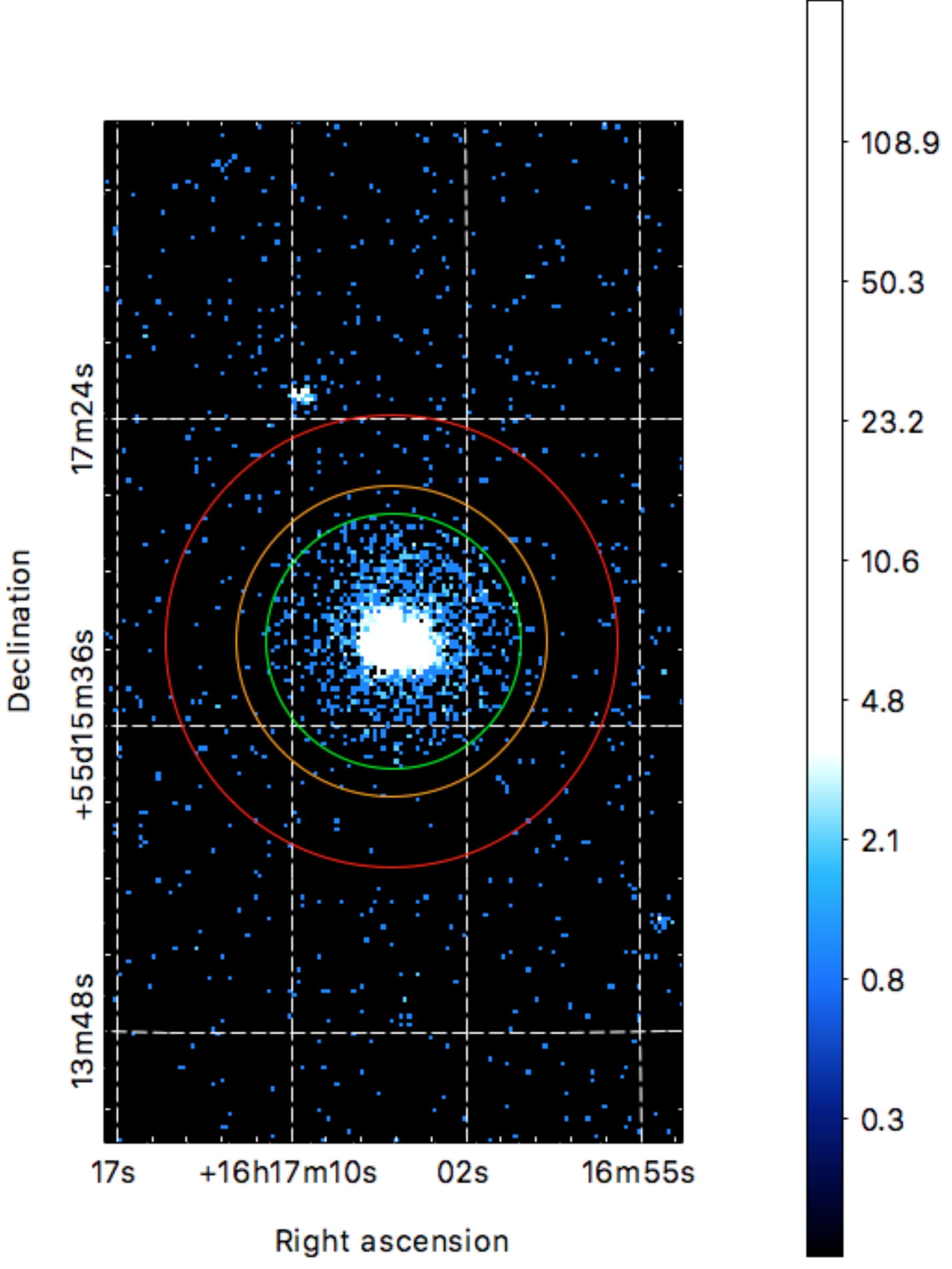}
\caption{Image (in counts) from a 10-second frame in the NUV using \texttt{gMap}.  This frame is selected from the image cube to define our aperture because it contains the peak of the largest flare on CR Draconis.  Apertures are represented by the colored circles, and correspond to 45, 55, and 80 arcseconds, respectively.  The color map is in a log scale, and is stretched to show the fainter wings of the source.  For reference, the maximum pixel contains 234 counts.\label{crdracoadd}}
\end{figure}

\edit1{The parameter \texttt{skyrange}, specified in degrees, tells \texttt{gMap} to make an image that's 6x6 arcminutes.  Figure \ref{crdracoadd} shows a frame during one of the larger flares from the 10-second image cube that we use to define our apertures.  With our apertures defined, we are now ready to create a light curve to examine the flares using \texttt{gAperture}:}

\begin{verbatim}
lc_data = gPhoton.gAperture(band='NUV',
                     skypos=[244.27247,
                             55.268069],
                     stepsz=10.,
                     csvfile='nuv_lc.csv',
                     radius=0.0045,
                     annulus=[0.0050,
                              0.0060])
\end{verbatim}

\edit1{This example is for the NUV band, a similar call can be used to make the FUV band light curve.  The parameter \texttt{radius} defines the photometric aperture in degrees, while \texttt{annulus} defines the inner and outer radii to use for background correction, also specified in degrees.  Since CR Dra has a significant proper motion, the \texttt{skypos} coordinates have been adjusted from the \texttt{gFind} and \texttt{gMap} commands to better match the epoch of the GALEX observations based on the \texttt{gMap} images.}

\begin{figure}[h!]
\includegraphics[width=0.46\textwidth,keepaspectratio]{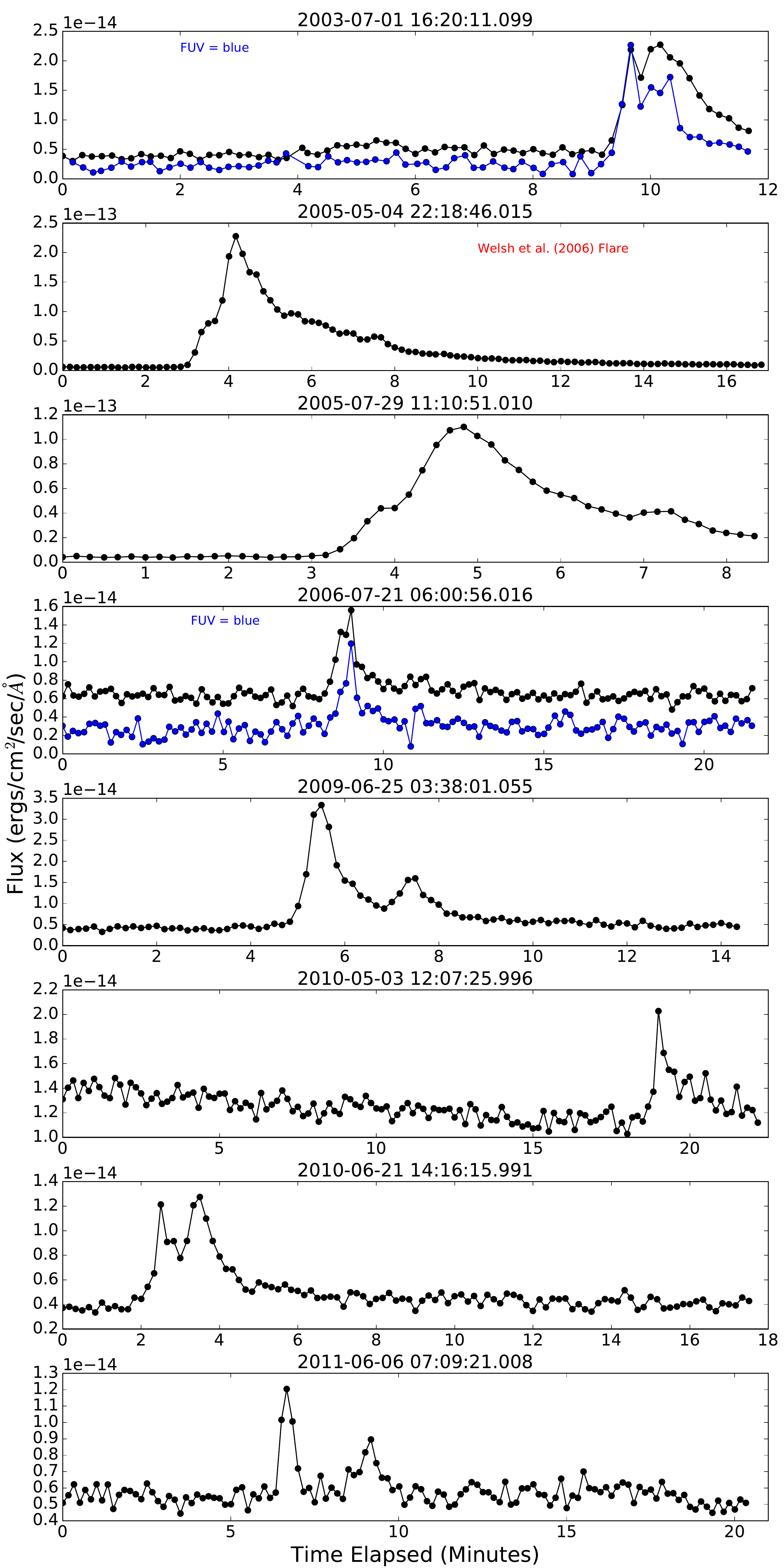}
\caption{Flares detected on CR Draconis using gPhoton, across the lifetime of the mission with a 10 second cadence. When available, FUV light curves are plotted (in blue) along with the NUV curves (in black). Fluxes have not been aperture corrected. Note that some of the larger flares reach sufficient brightness to exceed the non-linearity and saturation threshold of the detector; such points can be identified by checking whether bit 4 (''nonlinearity'') is set in the flag column returned by gAperture.
\label{crdraflares}}
\end{figure}

The largest observed flare in GALEX is the one reported in \citet{wel2006}; we have found seven additional flares, spanning from 2003 through 2011 and covering nearly 2 full orbits of the binary (Figure \ref{crdraflares}). When available, the FUV version of the light curves are shown in blue.  Several of the flares are double-peaked, and some show elevated levels of flux before or after the flare event.  There is also a range of amplitudes and durations, with some increasing in flux by less than a factor of two and lasting only a few minutes in duration. These short-duration flares have less energy than the longer duration, stronger flares, but also can occur more frequently, and thus may still impact the habitability of exoplanets in those systems \citep[e.g.,][]{ram2013}. There have been studies of the flare rates in resolved M dwarf binaries as a function of orbital separation, but the number of such binaries that have been observed for flares over their entire period range is small. CR Draconis is one candidate for such a system, however, and because the GALEX time baseline extends across two full orbital periods, gPhoton could help to improve the statistics over previous studies \citep{tam2008}. These are just two examples of how gPhoton allows researchers to characterize flares over timescales and energies that have been largely unexplored for stars other than the Sun.  In much the same way, we expect new discoveries when looking at short-term variability in pulsating stars, eclipsing systems, and extragalactic transients.

A more detailed astrophysical analysis of the flares \edit1{is beyond the scope of this introductory paper, which serves to present the database and software. In depth analysis of these and other stellar flares observable with gPhoton are reserved for future publications}. However, it is instructive to provide an outline demonstrating the basic work flow when creating the plot shown here. To define our photometric aperture, we used gMap to construct a deep coadd image, centered on CR Draconis, using all available photon events (Figure \ref{crdracoadd}). With our apertures defined, gAperture is used to construct comma separated value (CSV) light curve files of the target with 10 second time bins. We recommend bin sizes of between 10 and 30 seconds for first pass or exploratory analyses. We have found that variations over shorter timescales can exist (even in Figure \ref{crdraflares}) that may have astrophysical meaning. These can be detected at shorter time bins, even though the individual data points have larger uncertainty due to counting statistics. After the CSV light curve file is created, we wrote a separate script that reads in the CSV file, converts the $t_{\rm{mean}}$ timestamps from GALEX time to Julian Date, and then defined x-axis boundaries to center on each of the eight flare events of interest. Note that we did not apply aperture corrections to the fluxes shown in Figure \ref{crdraflares}, but such corrections are available in Figure\ 4 in \citet{mor2007}. An interpolation scheme is provided in gPhoton, using the values provided in \citet{mor2007}, called ``apcorrect1'' within the ``galextools.py'' module.

\section{Conclusion}
The gPhoton project extends the utility of the GALEX data set well beyond the scientific objectives of the original mission, most specifically towards the study of short time domain UV variability. Some of the techniques developed for gPhoton can be applied to other data sets produced by non-integrating detectors, particularly micro-channel plates. The fact that spatial analyses can be performed by making direct queries at the photon-level data, rather than artificially degrading the spatial resolution of the data by integrating and interpolating into pixelated images, offers potential advantages in terms of both the flexibility of the data archive and the computational overhead for some types of analysis. While not trivial, the corresponding data management and volume issues associated with storing and retrieving massive amounts of photon-level data are entirely solvable with appropriate use of existing, off-the-shelf database and storage technology. The behavior of the GALEX detector during very short timespans (which correspond to small spatial sampling of the detector) is not well characterized, and further work on improving the resolution of the detector flat fields, as well as correctly propagating flux uncertainties, will be required to derive the maximum utility from the photon-level data.

The gPhoton project is also a trial in an emerging paradigm for data archiving, where the functioning machinery for generating higher level data from lower---the calibration pipeline---is incorporated into the data archive itself. Even when preparation of the higher level data for archiving is well documented and comprehensible to future researchers, the priorities, interests, and needs of those users may not be the same as the data creators or archivists. At present, the standard recourse in such cases is to go back to some minimally reduced version of the data and create new tools or procedures for reducing the data from scratch. This can be onerous, time consuming, or impossible depending on the type of data, the quality of the documentation, and the availability of members of the original project team to answer inevitable questions. Especially when the data record observations that are unique or would be difficult to reproduce---for example, of rare astrophysical events in wavelengths only detectable above the atmosphere---an inability to reanalyze the data diminishes the long term value of results. Incorporating a \emph{functioning} calibration pipeline into the archive significantly lowers the barrier for independent research groups to modify that machinery to produce new science that was not anticipated by the original project teams.

\section{Acknowledgments}
The work presented in this paper was supported by the Mikulski Archive for Space Telescopes (MAST), which is funded by the NASA Office of Space Science via grant NNX09AF08G and by other grants and contracts. STScI is operated by the Association of Universities for Research in Astronomy, Inc., under NASA contract NAS5-26555.  We thank the GALEX beta testers for their feedback and patience, particularly Raghvendra Sahai, William Adler, Dun Wang, Luciana Bianchi, and Clara Brasseur.  We also thank several former members of the GALEX mission team for providing access to and consultation on the mission data, software, and calibration, including Patrick Morrissey, Don Neill, Min Hubbard, Tim Conrow, Ted Wyder, Tom Barlow, and Karl Forster. \edit1{We thank the \edit2{referee, Hans Moritz G{\"u}nther,} for comments that strengthened both the paper and codebase.} This research made use of Astropy, a community-developed core Python package for Astronomy \citep{astropy}. This research has made use of the SIMBAD database, operated at CDS, Strasbourg, France \citep{simbad}.

\bibliography{gphoton}

\end{document}